\documentclass[12pt]{article}
\usepackage{amsmath}
\usepackage{amssymb}
\usepackage{epsfig}
\newcommand{\pom}{\mathbb{P}}
\newcommand{\om}{\omega}
\newcommand{\omp}{\omega_\mathrm{I\!P}}
\newcommand{\oms}{\omega_s}
\newcommand{\omb}{\bar{\omega}}
\newcommand{\gamb}{\bar{\gamma}}
\newcommand{\chip}{\chi^{\prime}(\bar{\gamma})}
\newcommand{\as}{\alpha_s}
\newcommand{\asb}{{\bar \alpha}_s}
\newcommand{\be}{\begin{equation}}
\newcommand{\ee}{\end{equation}}
\newcommand{\bea}{\begin{eqnarray}}
\newcommand{\eea}{\end{eqnarray}}
\newcommand{\CF}{{\cal F}}
\newcommand{\CG}{{\cal G}}
\newcommand{\tbar}{\bar t}
\newcommand{\order}[1]{\mathcal{O}\left(#1\right)}
\newcommand{\eff}{\mathrm{eff}}

\newcommand{\ie}{i.e.\ }
\newcommand{\eg}{e.g.\ }

\begin{document}

\begin{flushright}
  DESY--02--049\\
  DFF 385/04/02\\
  LPTHE--02--23\\
  April 2002
\end{flushright}
\begin{center}
\vspace{2.5cm}
{\Large \textbf{\textsf{Expanding running coupling effects in the hard Pomeron
}}}
\vspace{0.7cm}

{\large \textsf{M.~Ciafaloni$^{(a,b)}$,
D.~Colferai$^{(c)}$,
G.~P.~Salam$^{(d)}$
and A.~M.~Sta\'sto$^{(b,e)}$}} \\
\vspace*{0.7cm}
$^{(a)}$ {Dipartimento di Fisica, Universit\'a di Firenze,  50019 Sesto
  Fiorentino (FI), Italy} \\
\vskip 2mm
$^{(b)}$ {INFN Sezione di Firenze, 50019 Sesto
  Fiorentino (FI), Italy} \\
\vskip 2mm
{$^{(c)}$II.\ Institut f\"ur Theoretische Physik,
    Universit\"at Hamburg, Germany}\\
\vskip 2mm
$^{(d)}$ {LPTHE, Universit\'es Paris VI et Paris VII, Paris 75005, France} \\
\vskip 2mm
$^{(e)}${H. Niewodnicza\'nski Institute of Nuclear Physics,
 Krak\'ow, Poland} \\
\vskip 2cm
\end{center}
\thispagestyle{empty}
\begin{center} {\large \textsf{\textbf{Abstract}}} \end{center}
\begin{quote}
  We study QCD hard processes at scales of order $\mathbf{k}^2 \gg
  \Lambda^2$ in the limit in which the beta-function coefficient $b$
  is taken to be small, but $\asb(\mathbf{k}^2)$ is kept fixed. The 
  (nonperturbative) Pomeron is exponentially suppressed in this
  limit, making it possible to define purely perturbative
  high-energy Green's functions. 
  The hard Pomeron exponent acquires diffusion and running
  coupling corrections which can be expanded in the $b$ parameter and
  turn out to be dependent on the effective coupling $b\asb^2 Y$.  We
  provide a general setup for this $b$-expansion and we calculate
  the first few terms both analytically and numerically.
\end{quote}
\newpage

\section{Introduction}

High energy hard scattering has received considerable attention in
recent years, due to the increase with the squared center-of-mass
energy $s$ of the experimental cross section seen at HERA \cite{HERA}.
The analysis of high energy QCD \cite{LLBFKL} shows that, at leading
$\log s $ level in which $\alpha_s$ is considered to be frozen, the
cross section is power behaved, and its exponent $\oms = \asb \chi_m
\equiv {\alpha_s N_c \over \pi} \chi_m$ is provided by the saddle-point
$\chi_m$ of the characteristic function $\chi(\gamma)$ of the BFKL
kernel.

When higher order corrections are taken into account \cite{NLLBFKL},
the theoretical analysis changes conceptually, because of running
coupling effects and of higher loop contributions to the kernel. First
of all, $\oms = \asb(t) \chi_m$ becomes $t$-dependent (where $t=\ln
k^2/\Lambda^2$ and $k$ is the momentum of the hard probe) and is no longer 
related to the leading singularity
in the $\om$-plane. Furthermore, the fact that $\asb(t)$ becomes
larger at small values of $k^2 = {\cal O}(\Lambda^2)$ causes the
existence of bound states (isolated, or in some cases dense $\om$
singularities) so that the leading one, the Pomeron $\omp$, becomes
crucially dependent on the strong coupling region and thus on the soft
physics. The question then arises of what one can really compute and
measure in a perturbative QCD approach.

It is a common belief that the ``hard Pomeron'' exponent is actually
measurable in two-scale hard processes (for example the two jet
production in hadronic collisions \cite{MuellerNavelet}, the forward
jet/$\pi^0$ production in deep inelastic scattering \cite{FORWARD} or
the $\gamma^*(k) \gamma^*(k_0) \rightarrow X$ process \cite{GAMGAM})
in which for sufficiently large $k^2,k_0^2 \gg \Lambda^2$, the cross
section is roughly determined by the exponent $\oms$, evaluated at
some average scale $k k_0$. It is also believed, though, that
diffusion corrections \cite{KOVMUELLER,ABB,LEVIN} to this exponent are
to be expected, and that for sufficiently large $Y \equiv \ln \frac{s}{k
k_0}$ the asymptotic, nonperturbative Pomeron $\omp$ takes over.

The interplay of the hard Pomeron regime (large $k^2$'s) with the
Pomeron one (large $Y$) has been the subject of intensive studies
\cite{JK,COLKWIE}, mostly dealing with running coupling effects and
diffusion corrections. Recently, it has been pointed out
\cite{CCS1,CCSS2} for various BFKL-type models that the transition to
the Pomeron regime is more like a sudden tunneling phenomenon rather
than a slow diffusion process.  A key result of this analysis is that
 for $k\simeq k_0 \gg \Lambda$, the hard Pomeron behaviour dominates if
$\oms(kk_0) Y$ is large but not too large, its critical value being of
the order of $\bar{t}$, if $\omp \simeq \frac{\chi_m}{b\bar{t}}$
parametrises the size of the Pomeron exponent ($\tbar$ is the scale at 
which $\asb(t)$ is regularized). Thus, the puzzling
question arises again: is there really room for an observable hard
Pomeron regime?

Some source of optimism comes from the fact that the physical Pomeron
of soft physics is indeed rather weak (the exponent is around $0.08$
in most estimates \cite{DL}). Thus if $\omp$ is such a Pomeron, the
effective critical value of $\oms Y$ may be sufficiently large for the 
hard
Pomeron to be seen. It is true that, the physical Pomeron is a
complicated unitarity effect which turns out to be weak and
approximately factorized for well studied \cite{CAPELLA} but not well
understood reasons. However, the trend of subleading corrections
\cite{NLLBFKL} is that of reducing the high-energy exponents and the
diffusion coefficient, and this presumably slows down the onset of the
nonperturbative behaviour as well.

A different way of studying the problem, motivated in the present
paper, relies on a somewhat different, but crucial observation
(Sec.~2). The size of the tunneling probability at scale $t=\ln
k^2/\Lambda^2$ is easily seen to be of the order
$\exp[-\frac{1}{b\asb(t)} g(\asb(t))]$, where $b$ is the beta-function
coefficient which determines the running of the coupling $\asb(t) =
1/bt$, and $g(\asb(t))$ is a function whose exact form depends on the
nature of the problem being studied. Therefore, in the formal limit $b
\rightarrow 0$ with $\asb$ 
taken to be fixed there is no tunneling to the Pomeron and only the
hard Pomeron survives.  In this paper, we take advantage of the
Pomeron suppression for $b \rightarrow 0$ in order to study the
perturbative part of the gluon Green function by expanding it in the
$b$ parameter both analytically and numerically.  It turns out, that
in BFKL-type models $b$ plays the role of $\hbar$ in a semiclassical
expansion, and thus the above observation is actually the basis for a
full calculation of the Green's function exponent which allows to find
the diffusion corrections to the hard Pomeron in a systematic way
(Sec.~3).

We stress the point that we are expanding in $b$ 
the gluon Green function
with two hard scales ($t$,$t_0 \gg 1$) by keeping $\asb(t)$ and possibly 
$\asb(t_0)$ 
fixed, we are not expanding the renormalization group logarithms themselves.
Therefore, our expansion is somewhat different from previous related 
expansions of the Green's function \cite{KOVMUELLER,ABB,LEVIN} and from analogous expansions applied to the collinear factorization limit 
\cite{Thorne,ABF}.
We also stress that our expansion concerns the
perturbative part of the gluon Green's function, the Pomeron one being
exponentially suppressed as stated before. Here a technical point
arises: how to define such perturbative part? We argue in Sec.3 that
the customary \cite{JK} double $\gamma$ representation can be given a
precise perturbative meaning in a general BFKL-type model, apart from
the intrinsic renormalon ambiguities, which come from the Landau-pole
regularisation and are suppressed like $\sim
\exp(-\frac{1}{b\asb(t)})$. We thus overcome the difficulties found in
Ref.~\cite{ABB} for the principal value prescription, by a careful
choice of the integration contours.

The $b$-expansion provides a hierarchy of diffusion corrections
to the Green's function exponent for $t\simeq t_0$ depending on the
parameter $z=\frac{1}{2} b^2 \asb^2 \chi_m \chi_m^{''} \asb^2 Y^2$,
introduced in Ref.~\cite{CMT}, and which turns out to be of the order
$b^2$. For $z \ll 1$ or $\oms Y \ll t$ we recover the well known
\cite{KOVMUELLER,ABB,LEVIN} $\asb^5 Y^3$ terms, and we compute new
$\sim \asb^4 Y^2$ and $\asb^3 Y$ diffusion corrections which change
the normalisation and the exponent at order $b^2$ (Sec.3). The
estimate of the $Y^3$ and $Y^2$ terms is confirmed by the numerical
methods (Sec.5), which could be extended to the realistic model
\cite{CCSS3} with the purpose of resumming subleading effects,
 as in the improved equation \cite{CCS2},
 or in the alternative approaches 
\cite{ABF2000,BFKLP,SCHMIDT,FRSV,FADIN2000}.
  We are also able to evaluate the convergence radius
$z_c$ of the above series, and the large $z$ behaviour which turns out
to be oscillatory \cite{CMT} and is actually masked by the onset of
the asymptotic Pomeron. An amusing result is an essentially explicit
evaluation of the perturbative Green's function for the Airy diffusion
model (Sec.4) which improves the estimates of Ref.~\cite{CMT}.
 
Another application of the $b$-expansion is in the numerical
determination of high-energy Green's functions. Current practice in
numerical solutions of the small-$x$ evolution equations (see for
example \cite{Kwiecinski:1999hv,LUND}) is to evaluate a perturbative result
with an infrared regularised coupling (\eg with a cutoff) and then
define the perturbatively accessible region in $Y$ as that insensitive
to the choice of regularisation.  While physically intuitive, this
approach has two significant disadvantages.  Firstly the choice of a
particular set of regularisation schemes is quite subjective. Secondly
the formal limit on the perturbative region is determined by
tunneling, \ie $Y_\mathrm{max} \sim 1/\as(t_0)$, whereas perturbative
predictions can in principle be made up to values of $Y$ of order
$1/\as^2(t_0)$. As is discussed in section~\ref{sec:num}, the
$b$-expansion offers a solution to both these problems.

The overall picture emerging from this work is that the hard Pomeron
and its corrections can be properly defined in the $b$-expansion,
and its features can be found either numerically or analytically at
semiclassical level in a precise way.  Beyond this level of accuracy
some ill-defined exponentially suppressed terms come in which are
nonperturbative in both $b$ and $\asb$ and among them the asymptotic
Pomeron itself. It is not known whether the size of such contributions
may be hinted at or restricted on the basis of perturbative arguments.

\section{Hard Pomeron vs Pomeron in the perturbative regime}

The basic problem of small $x$ QCD is to determine the gluon Green's
function $G(Y;t,t_0)$, where the logarithmic variables $Y=\log \,{
  s/kk_0}$, $t = \log \, {\mathbf k}^2\!/\Lambda^2$ and $t_0 = \log
\,{\mathbf k_0}^2\!/\Lambda^2 $ are related to the center-of-mass
energy $\sqrt{s}$ and to the transverse momenta of the gluons
$\mathbf{k}$, $\mathbf{k_0}$ ($k=|\mathbf{k}|$, $k_0=|\mathbf{k_0}|$) and 
$\Lambda$ is a QCD parameter. In
BFKL-type models, the function $G$ satisfies the following evolution
equation 
\begin{eqnarray}
\frac{\partial}{\partial Y} \; G(Y;t,t_0)  & = &  K \otimes G \nonumber \\
G(0; t,t_0) & = & \delta(t-t_0) \; ,
\label{bfklll}
\end{eqnarray} 
where $K$ is a kernel in $t$-space. It is well known \cite{LLBFKL}
that, at leading level in $\log s $, the kernel $K(t,t^{\prime}) =
\asb K_0(t-t')$, is scale invariant with symmetric characteristic
function $\asb \chi_0(\gamma) = \asb \chi_0(1-\gamma)$, so that the
high-energy exponent of Eq.~(\ref{bfklll}) becomes
\begin{equation}
  \omp \; = \; \oms^0 \; = \; \asb \max_{\sigma \in \mathbb{R}} \; 
\chi_0(\frac{1}{2} + i
  \sigma) \; = \; \asb \chi_0(\frac{1}{2}), \hspace*{1cm} \asb =
  \frac{N_c \alpha_s}{\pi}.
\label{omegasll} 
\end{equation} 
Therefore, in the LLA this exponent is both the leading singularity in the
$\omega$ plane (conjugated to $Y$) and the saddle point value of the
characteristic function $\chi$. For not too large values of $t-t_0$, a
good approximation of $G$ is provided by the diffusion model
\begin{equation} G(Y;t,t_0) \; \simeq \; \frac{1}{\sqrt{4 \pi D^0
      \oms^0 Y } } \exp\bigg(\oms^0 Y - \frac{(t-t_0)^2}{4 D^0 \oms^0
    Y} \bigg), \hspace*{1cm} D^0=\frac{1}{2} \frac{\chi_0^{\prime
      \prime}(1/2)}{\chi_0(1/2)} \; .
\label{soldiff}
\end{equation} 
When subleading corrections in $\log s$ are evaluated \cite{NLLBFKL}
running coupling and higher loop corrections come into play.  The
kernel $K$ is no longer scale invariant --- being dependent on $\asb(t)$
in a essential way --- and may have, besides the continuum spectrum, a discrete
one whose maximum eigenvalue is $\omp$.  Furthermore, the fact that
$\asb(t)$ becomes stronger at small values of $k^2 \simeq {\cal
  O}(\Lambda^2)$ favours the diffusion towards small $t$ values, so
that eventually the asymptotic exponent in Y, that is $\omp$, is
crucially dependent on the soft physics. The magnitude of $\omp$ will
then depend on the details of the physical regularisation of $\asb(t)$
adopted in the region around the Landau pole.
We shall refer for definiteness to kernels of the slightly asymmetrical form%
\footnote{This form is a standard one for a solution based on the
  $\gamma$-representation (Mellin transform), and can be shown to be 
sufficiently
  general with the approach of the $\omega$-expansion \cite{CCS2} in
  which $\tilde{K}$ is taken to be dependent on the variable $\omega$,
  conjugated to $Y$. }
\begin{equation}
\label{eq:AsymKernel}
K = \asb(t) \tilde{K}(t-t')
\end{equation}
where $\asb(t)$ embodies the strong coupling boundary condition
below some scale $\tbar$ and $\tilde{K}$ is scale invariant.

On the other hand it is still believed that, when $t$ and $t_0$ are
fixed by the experimental probes to be sufficiently large (for example
in scattering of two highly virtual photons) the diffusion to low
$t's$ is suppressed and the $s$ dependence is basically power-like
and $t$-dependent
\begin{equation}
G(Y; t,t_0) \; \simeq \; \frac{1}{\sqrt{4\pi D \oms(\frac{t+t_0}{2}) Y}} \exp\bigg(
\oms(\frac{t+t_0}{2}) Y + {\rm corrections } \bigg) \; ,
\hspace{0.5cm} Y \gtrsim \; t,t_0 \;  \gg 1 \; ,
\label{solpert} 
\end{equation}
where the exponent $\oms(t)$ is now perturbatively calculable.  It
should be noted however that $\oms(t)$ is no longer a singularity in
the $\om$ plane in that case.

Some insight into the interplay of $\oms(t)$ and $\omp$ is gained
\cite{CMT} in simplified models like the Airy diffusion model
\cite{AIRY} or collinear models \cite{CCS1} in which the gluon Green's
function in $\om$ space is factorized  for  a kernel $K$  of the 
form (\ref{eq:AsymKernel}) as follows
\begin{equation}
{\cal G}_{\om}(t,t_0) \; = \; \frac{t_0}{\omega}
\left({\CF_{\om}}(t) \tilde{\CF}_{\om}(t_0)
\Theta(t-t_0) \; +  \; t \longleftrightarrow t_0 \right) \; ,
\label{greenfactor}
\end{equation}
where ${\cal F}_{\om}(t)$ is the (right) regular solution for $t \rightarrow
\infty$ of the stationary homogeneous diffusion equation $(\om - 
  K){\cal F} =0 $, and $t_0 \tilde{\cal F}(t_0)$ is instead the (left) regular one  for
$t_0 \rightarrow -\infty$ ($k_0^2 \rightarrow 0$).  In such models the
strong coupling boundary condition on $\tilde{\CF}_{\om}$ determines
it in the form
\begin{equation}
\tilde{\cal F}_{\om}(t_0) \; = \; {\cal F}_{\om}^I(t_0) \; + \; S(\om) {\cal F}_{\om}(t_0),
\label{fdec}
\end{equation}
where ${\cal F}^I$ is now an irregular solution for $t_0 \rightarrow
\infty$ and the reflection coefficient $S(\om)$ carries the leading
spectral singularities of the problem, in particular the Pomeron
singularity $\omp$.  By using the decomposition (\ref{fdec}) of
$\tilde{\cal{F}}$ into its perturbative and nonperturbative components
in representation (\ref{greenfactor}) we obtain the following general
expression for the gluon Green's function
\begin{equation}
G(Y; t,t_0) \; = \; \int \frac{d \om}{ 2 \pi i} e^{\om Y} \, \frac{t_0}{\omega} \bigg[
\CF_{\om}(t) \CF_{\om}^I(t_0) +  
S(\om) \CF_{\om}(t) \CF_{\om}(t_0) \bigg]\;, \hspace{0.5cm} t>t_0 \; .
\label{greendec1}
\end{equation}
This expression is interpreted as a decomposition into perturbative
and nonperturbative part of the gluon Green's function
\begin{equation}
G(Y;t,t_0) \; = \; G_{pert}(Y;t,t_0) \; + \; G_{\pom}(Y;t,t_0) \; .
\label{greendec2}
\end{equation}
The second term on the right hand side of Eqs.~(\ref{greendec1}) and
(\ref{greendec2}) defines the Pomeron contribution at $Y \rightarrow
\infty$
\begin{equation}
G_{\pom}(Y;t,t_0) \; \simeq \; R \exp(\omp Y) \; {\cal F}_{\omp}(t) \;
{\cal F}_{\omp}(t_0) \frac{t_0}{\omp} \; ,
\label{greennonp}
\end{equation}
where we set $R={\rm Res}_{\omp} S(\om)$. For instance, in the case of
the Airy diffusion model with cutoff at $t=\tbar$, $R$ is roughly
estimated \cite{CCSS2} to be equal $R=\Delta \omp/\xi_0
[\mathrm{Ai}'(-\xi_0)]^2$, on the basis of the bound state condition
$\mathrm{Ai}(-\xi_0)=0$, with \mbox{$\Delta \omp 
=\oms(\tbar)-\omp\simeq$} \\
$ \simeq [\sqrt{D} b\asb(\tbar)]^{2/3} \omp$, where 
$D=\chi_m''/2\chi_m$.

 Result (\ref{greennonp})
suggests an estimate of the Pomeron contribution on the basis of the
$\gamma$ representation of the regular solution
\begin{equation}
{\cal F}_{\om}(t) \; =\; \int \frac{d\gamma}{2 \pi i} e^{(\gamma-1/2)
  t - \frac{1}{b\om} X(\gamma)} \; ,
\label{gammarep}
\end{equation}
where we have used
\begin{equation}
\asb(t) = \frac{1}{b t} \; , \hspace{2cm} b=11/12 \; .
\label{alphastrong}
\end{equation}
This representation was argued to be valid \cite{CCS2} in the
perturbative regime $t,(b\om)^{-1} \gg 1$ for kernels of type
(\ref{eq:AsymKernel}) for which $X^{\prime}(\gamma) = \chi(\gamma)$ is
the characteristic function of $\tilde{K}$, and can be extended to the case of 
more general kernels in the framework of $\omega$-expansion.
Suppose now, that besides $b\asb(t) \simeq \frac{1}{t}$, $b\omp$ also
is a small parameter. This is possible in a realistic case (that is $b
\sim {\cal O}(1)$), if, for some reason $\omp$ itself is small (weak
Pomeron) and is also possible in the case of fixed $\omp$ in the limit
of $b \rightarrow 0$, i.e. in the small $b$-expansion being considered
here. Then, one can estimate the magnitude of ${\cal F}_{\omp}$ from
the saddle point condition applied to (\ref{gammarep})
\begin{equation}
b  \omp  t \; = \; \chi(\frac{1}{2}-p)
\label{saddleomp}
\end{equation}
which implicitly defines $p(b \omp t)$ and yields
\begin{equation}
{\cal F}_{\omp}(t) \; \simeq \; \exp\bigg(-\frac{1}{b  \omp}
\int_{\chi_m}^{b \omp t} \! \!  p(x) dx \bigg) \sqrt{\frac{b \omp }{-2
    \pi \chi^{\prime}[p(b \omp t)] } }\; , \hspace*{0.5cm}
\chi_m=\chi(1/2),
\end{equation}
where we have used the relation 
\begin{equation}
p(b \omp t)\, b t + \frac{1}{\omp} X(\frac{1}{2} - p(b \omp t)) =
\frac{1}{\omp} \int_{\chi_m}^{b \omp t} \! \! p(x) dx \; .
\label{anomrel}
\end{equation}

Because of the decomposition (\ref{greendec2}) and of the  perturbative
behaviour (\ref{solpert}) we obtain the estimate
\begin{equation}
\frac{G_{\pom}(Y;t,t)}{G_{pert}(Y;t,t)} \; \simeq \; 
\frac{ b t R}{|\chip|\omp}\sqrt{\frac{D\oms Y}{\pi}}\exp \bigg[ 
(\omp-\oms(t)) Y - \frac{2 }{b \omp } \int_{\chi_m}^{b \omp t}\!
\! p(x) dx \bigg] \; .
\label{pomtopert}
\end{equation}

In other words, the Pomeron is potentially leading for $Y \rightarrow
\infty$ because $\omp > \oms(t)$ for $t \gg 1$, but is exponentially
suppressed with the hard scale parameter $t$. We notice that the
suppression exponent is dependent on the parameter $b\omp t = \omp /
\asb$, and also on the model being considered (implying a different
solution for $p(x)$ from (\ref{saddleomp})), but is always of the
form
\begin{equation}
\frac{1}{b \asb(t)} g(\asb(t)),
\label{suprfac}
\end{equation}
(the $\omp$ dependence is implicit in $g$).
 For instance in the case of the Airy diffusion model we get 
\begin{equation}
x = \chi_m [ 1 + D p(x)^2] \; \; \longrightarrow \; \; p(x) =
\frac{1}{\sqrt{D}} \sqrt{\frac{x}{\chi_m}-1} \; .
\end{equation}
Therefore in the limit $b \omp t  \gg 1$ we get the suppression factor
\begin{equation}
\exp\bigg[-\frac{4}{3 b \asb} \sqrt{\frac{\omp}{D \oms(t)} }\, \bigg]
\sim \exp  [-{\rm const} \; t^{3/2}  ] \; ,
\label{airysupr}
\end{equation}
with the function $g \sim {\rm const}/\sqrt{\asb}$.  On the other hand
in the collinear model with ${\chi(\gamma) = 1/(\gamma(1-\gamma))}$
one obtains
\begin{equation}
x = \frac{1}{\frac{1}{4}-p(x)^2} \; \;  \longrightarrow \; \; p(x) =
\sqrt{\frac{1}{4}-x} \; ,
\end{equation} 
which yields the suppression factor of the form,
\begin{equation}
\exp[-\frac{1}{b\asb} ] \simeq \exp(-t), \; \; g \sim {\rm const} \; .
\label{collsupr}
\end{equation}

We conclude that, at large values of $t$, the transition probability
to the Pomeron behaviour is typical of a tunneling phenomenon and is
of the general form
\begin{equation}
\exp \bigg(-\frac{1}{b\asb}g(\asb) \bigg) \; .
\label{tunnelfactor}
\end{equation}
so that the $b$ parameter at fixed values of $\asb$ plays a role of
$\hbar$ in a semiclassical approach in quantum mechanics.

A better insight into the tunneling phenomenon is gained in the simple
example of the Airy diffusion model which is obtained by expanding the
kernel eigenvalue into the second order in $\gamma$. By using this
approximation, the evolution equation (\ref{bfklll}) takes the form
\begin{equation}
\frac{\partial }{\partial Y} G(Y;t,t_0) \; = \; \oms(t) \bigg[ 1 \; +
\; D \frac{\partial^2 }{\partial t^2}  \bigg] G(Y;t,t_0) .
\label{airyequation}
\end{equation}
We make then the following change of variables
\begin{equation}
x \; = \; b \int^t \! \frac{dt }{\sqrt{\oms(t) D }} \; , 
\label{xvar}
\end{equation}
which in this case results in
\begin{equation}
x \; = \; \frac{2}{3} \frac{(b t)^{3/2}}{\sqrt{D \chi_m}} \; .
\label{xvar1}
\end{equation}
Making use of (\ref{xvar},\ref{xvar1}) and performing additionally the
transformation \mbox{$G = x^{-\frac{1}{6}} \hat{G}$} one can recast
Eq.~(\ref{airyequation}) into the Schroedinger-like diffusion
equation
\begin{equation}
\label{schroedinger}
-\frac{\partial}{\partial Y} \hat{G} \; = \; \bigg[  \hat{p}^2  \; +
\; V(x) \bigg] \hat{G} \; ,
\end{equation}
where now the rapidity $Y$ is interpreted as the imaginary time and
$\hat{p} \equiv {ib } \frac{\partial}{\partial x}$ is the ``momentum''
variable conjugated to $x$. The potential $V(x)$ has the following
form
\begin{equation}
V(x) \; = \; -\oms[t(x)] - \frac{5}{36} \, b^2 \, \frac{1}{x^2} \, ,
\end{equation}
and is basically provided by the $-\oms[t(x)]$ part dependent on the
running coupling $\asb(t)$, apart from the quantum corrections
proportional to $b^2$.
 
\begin{figure}[t]  
  \vspace*{0.0cm}  
     \centerline{  
         \epsfig{figure=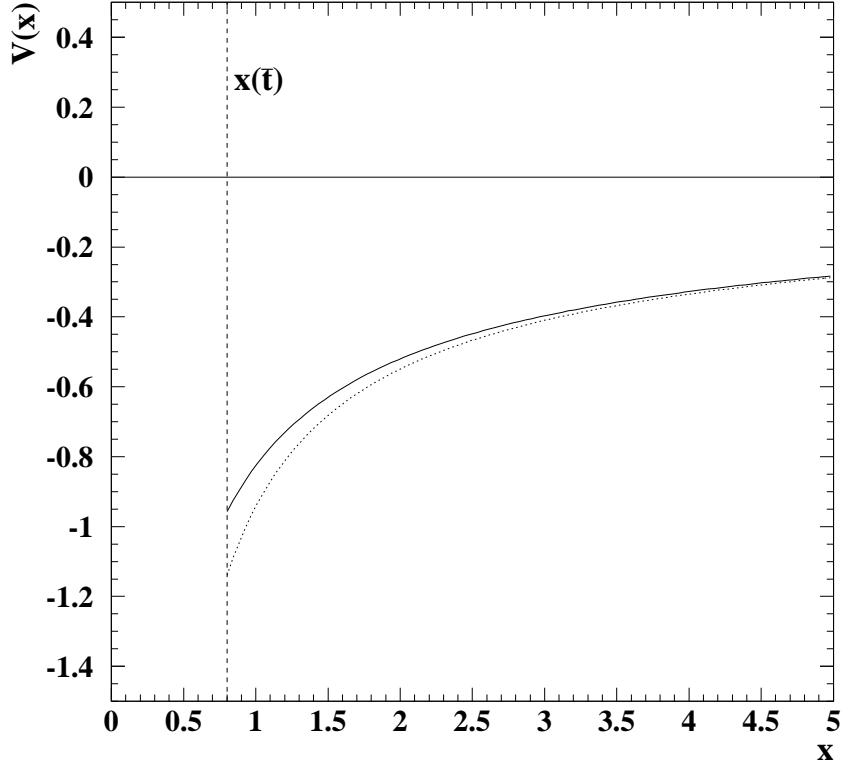,width=12cm}  
           }  
\vspace*{0.5cm}  
\caption{\it Shape of the potential $V(x)$ in Schroedinger-like equation
  from the diffusion model with running coupling,
  Eq.~(\ref{schroedinger}).  Solid line: leading term $-\oms(t(x))$;
  dotted line:  leading term + quantum corrections
  $-\frac{5}{36}\frac{b^2}{x^2}$.  The dashed line illustrates the
  boundary $x(\bar{t})$ at which one has specified the regularisation
  of $\asb(t)$.
\label{fig:potential}}  
\end{figure}

In this interpretation, which differs slightly from the
$\omega$-dependent ones discussed previously \cite{CCS1}, the
non-perturbative 
parameter $\omp$ is directly given by the lowest energy state in the
stationary equation corresponding to (\ref{schroedinger}), whose
potential is pictured in Fig.~\ref{fig:potential}. Therefore the
barrier is directly that arising at energy $-\omp$ by the fact that
$\oms(t)$ decreases to zero at large values of $t(x)$.  Both the
binding of the Pomeron and its suppression at large values of $t$ are
provided by the running of $\asb(t)$ which implies that the strong
coupling region is much more ``attractive'' than the weak coupling
one. The detailed estimate of $\omp$ depends crucially on the shape of
the potential well for small values of $t$ which instead is provided
by specifying the strong coupling behaviour of $\oms(t)$ and
$\asb(t)$. For instance in the case of the cut-off prescription
(Fig.~\ref{fig:potential}) 
\begin{equation}
\asb(t) = \Theta(t-\bar{t}) \frac{1}{bt} \, ,
\label{cutoffasb}
\end{equation}
one requires the solution to vanish at the boundary $\bar{t}$,
\begin{equation}
\tilde{\CF}_{\omp}(\bar{t}) = 0 \, .
\label{cutoffsol}
\end{equation}
 The suppression
factor on the other hand is basically provided by the perturbative regime
and  by Eq.~(\ref{pomtopert}) in the semiclassical approximation.

\begin{figure}[t]  
  \vspace*{0.0cm}  
     \centerline{  
         \epsfig{figure=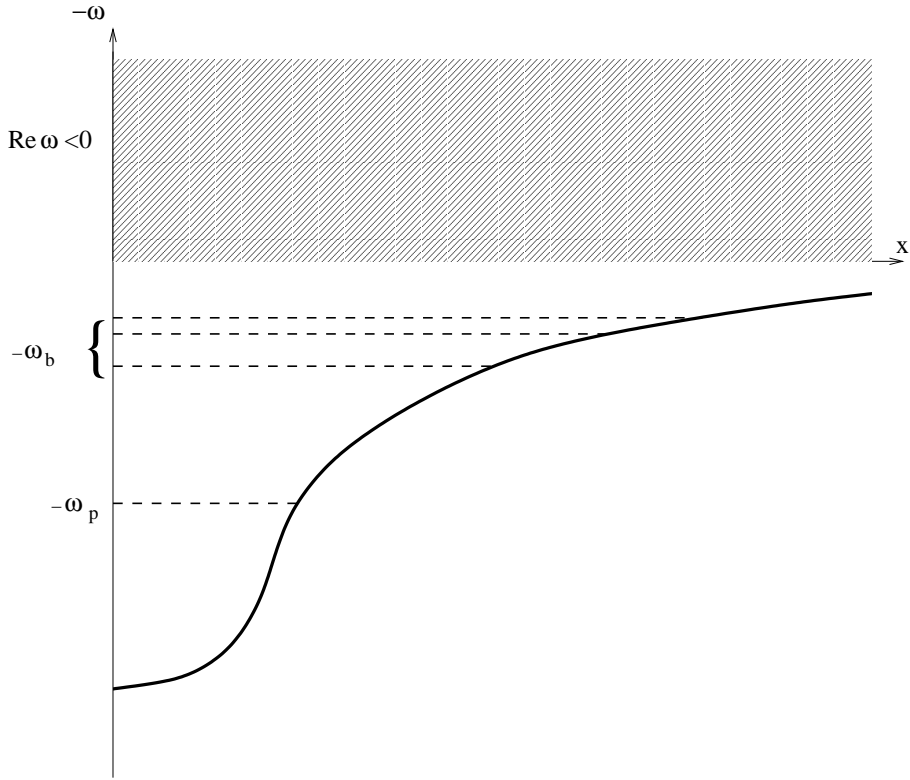,width=12cm}  
           }  
\vspace*{0.5cm}  
\caption{\it Schematic representation of the states in an arbitrary
  potential well. The continuum, which corresponds to \mbox{Re $\om<0$}, is
  the perturbative part; $\omp$ is the lowest lying bound state
  corresponding to the Pomeron; $\om_b$ denotes the subleading (discrete 
or
  continuum) bound states.
\label{fig:states}}  
\end{figure}

In the potential-like interpretation above, the Pomeron contribution
to the Green's function corresponds to the lowest lying bound state
and nearby ones (which are solutions to the homogenous equation) and
the perturbative contribution roughly corresponds to the continuum of
positive energies, yielding singularities at \mbox{Re$ \, \om \le 0$},
see Fig.\ref{fig:states}.  This decomposition is however not unique
(neither was (\ref{fdec}), because of ambiguities in the definition of
the irregular solution $\CF^I_{\om}$).  In fact, subleading bound
states, $\om_b < \om_{\pom}$, see Fig.~\ref{fig:states} can be
incorporated into the perturbative part also, depending on the size of
$\oms(t)$ . Since they are still suppressed by
$\exp(-\frac{1}{b\asb(t)})$ factors, like the Pomeron, they appear to
play a role similar to renormalon ambiguities in QCD perturbation
theory.  The important point is that the whole Green's function is
well defined, given the strong coupling boundary condition, as is also
the Pomeron contribution, provided by the lowest energy bound state
$-\omp$ in the potential $V(x)$.
\section{Perturbative Green's function and its $b$-expansion}

We have just understood on the basis of a few solvable examples, that
the gluon Green's function can be decomposed (in a non-unique way)
into a perturbative part, which gives the $\oms(t)$ exponent,
Eq.~(\ref{solpert}) at large values of $t$, and a Pomeron part which
is leading for $Y \rightarrow \infty$, but is suppressed by the
tunneling factor (\ref{pomtopert}) at large $t$'s.  We shall now
analyse a similar decomposition in the more general case of a BFKL
model with running coupling, in which $K$ has the form (\ref{eq:AsymKernel})
where $\tilde{K}$ is a scale invariant kernel with eigenvalue function
$\chi(\gamma)$. It is known \cite{JK} that for such a model, a formal
solution for the Green's function in the perturbative regime is
provided by the double $\gamma$-representation
\begin{equation}
{\cal G}_{\om}(t,t_0) \; = \; \int^{c+ i\infty}_{c-i \infty} \! \!
\frac{d\gamma}{2\pi i} \int_{\gamma}^{c(\infty)} \!\frac{d\gamma_0 \,
  t_0}{\om} \; \exp \bigg[
(\gamma- \frac{1}{2})t-({\gamma}_0- \frac{1}{2}) t_0 - \frac{1}{b \om}
X(\gamma)+\frac{1}{b \om} X(\gamma_0) \bigg]
\label{doublegammarep}
\end{equation}

This expression should be used with care for several reasons. Firstly
it is a formal solution of Eq.~(\ref{bfklll}) with running coupling
$\asb(t)=1/bt$ which diverges at $t=0$ (Landau pole). Thus it may be
acceptable for $t,t_0 \gg \bar{t} $ only, where $\bar{t}$ parametrizes
the boundary of the strong coupling region, when $\asb(t)$ departs
from the perturbative expression $1/bt$. Furthermore, the
representation (\ref{doublegammarep}) has an essential singularity at
$\om=0$. Therefore, the Pomeron singularity $\omp>0$ arising from the
regularisation of the strong coupling is not obviously included in
this representation. Finally, the definition (\ref{doublegammarep}) is
ambiguous also, because the convergence path of the $\gamma_0$ contour
is not specified for $\gamma_0 \rightarrow \infty$. Any further
specification will represent the addition of a single
$\gamma$-representation (\ref{gammarep}), that is a solution of the
homogenous 
equation (\ref{bfklll}) (as is the Pomeron).  Given these questions
we shall first provide a more rigorous definition of the perturbative
part of the gluon Green's function.
\subsection{Relation of the double-$\gamma$ representation to the spectrum}

We start considering the double $\gamma$ representation
(\ref{doublegammarep}) in the context of the principal value
regularisation of the Landau pole in Eq.~(\ref{eq:AsymKernel}). Such
regularisation was already considered in Ref.~\cite{ABB} and was shown
to result in a continuous $\omega$ spectrum, with $\omega$ having all
real values from $-\infty$ to $+\infty$. This seems paradoxical at
first sight, but is actually hardly surprising, since in such a case
the strong coupling constant $\asb(t)$ is not positive definite and is
unbounded.
Since we actually aim at a physical Landau pole regularisation in
which $\asb(t)$ keeps its positivity and is bounded, we should
properly handle in  Eq.~(\ref{doublegammarep}) the contribution of the
positive end of the $\omega$-spectrum, which is an artifact of the
principal value regularisation.
 
Let us recall \cite{ABB} that, in the analytical principal value
regularisation\footnote{Other prescriptions may lead to different results. For instance
  taking a sharp \mbox{cutoff} \mbox{$P(t) = \frac{1}{t}\Theta(|t|-\epsilon)$} separates the
  Hilbert space into the positive-$t$ and negative-$t$ eigenfunctions, with a distortion
  which, for large $t$ values, is
  again $\order{\exp(-\frac{1}{b\asb(t)})}$.}
\begin{equation}
 P\left(\frac{1}{t}\right) = \frac{1}{2} \left(\frac{1}{t+i\epsilon}+\frac{1}{t-i\epsilon}\right)
\end{equation}
 the
eigenfunctions in $\gamma=\frac{1}{2}+i \nu$ space of the kernel (\ref{eq:AsymKernel}) take
the form
\begin{eqnarray}
\CF_{\om}(\gamma) & = & \exp[ -\frac{1}{b\om} X(\gamma)]\; ,
\nonumber \\
\bar{\CF}_{\om}(\gamma_0) & = & -\frac{\chi(\gamma_0)}{2\pi
  b\om^2}\exp[ \frac{1}{b\om} X(\gamma_0)] \; ,
\label{eigenfungam}
\end{eqnarray}
where we have used the representation $t=-\partial_{\gamma}$ to derive the
(right) eigenfunctions, and the
conjugate (left) eigenfunctions due to the asymmetrical role of the coupling
in Eq.~(\ref{eq:AsymKernel}), are defined by
$\bar{\CF}_{\omega}(\gamma_0)=-\frac{\partial_{\gamma_0}}{2\pi\omega}
\CF^*_{\omega}(\gamma_0)$ and satisfy the delta-function normalisation
$(\bar{\CF}_{\omega},\CF_{\omega'})=\delta(\omega-\omega')$. As a
consequence, the Green's function $\CG_{\omega}=(\omega-K)^{-1}$
satisfies the spectral representation
\begin{equation}
\CG_{\om}(\gamma,\gamma_0) \; = \; \int_{-\infty}^{+\infty} \!
\frac{d\om'}{\om-\om'} \,\CF_{\om'}(\gamma) \,
\bar{\CF}_{\om'}(\gamma_0) \; ,
\label{greenspectral}
\end{equation}
where the variables $\gamma=1/2+i\nu$ and $\gamma_0=1/2+i\nu_0$ run
along the \mbox{Re $\gamma = 1/2$} axis.  In order to evaluate the
integral (\ref{greenspectral}) we shall distort the $\om'$ contour so
as to go to \mbox{Im $\om'>0$} (\mbox{Im $\om'<0$}) for $\nu_0>\nu$
($\nu>\nu_0$).  We have to additionally assume  that $\chi_m<0$,
so that $i(X(\gamma_0)-X(\gamma))$ has always the sign of $\nu_0-\nu$.
The case $\chi_m>0$ will then be treated by analytical continuation.
The result of this contour integration in Eq.~(\ref{greenspectral})
reads then
\begin{equation}
\CG_{\om}(\gamma,\gamma_0) \; = \; -\frac{1}{\omega}\frac{\partial}{\partial \gamma_0}
\bigg\{\epsilon_{\om}(\nu_0-\nu) \exp \bigg[\frac{1}{b\om}
(X(\gamma_0)-X(\gamma)) \bigg] \bigg\}\; ,
\label{greenprincipal}
\end{equation}
with
\begin{equation}
\epsilon_{\om}(\nu_0-\nu) = \left\{ \begin{array}{rl}
\Theta(\nu_0-\nu), & \mbox{Im $\om >0$}   \\
-\Theta(\nu-\nu_0), & \mbox{Im $\om <0$} \; .
\end{array}
\right.
\end{equation}
Then, by performing a Fourier transform to $t$-space, 
it is  straightforward to derive from the expression
(\ref{greenprincipal}) the double $\gamma$-representation of
Eq.~(\ref{greenpertgam}) with the additional constraint that the contour
$c(\infty)$ is chosen so as to make \mbox{Re $(\gamma_0-\gamma)/\om >
  0$}.

\begin{figure}[t]  
  \vspace*{0.0cm}  
     \centerline{  
         \epsfig{figure=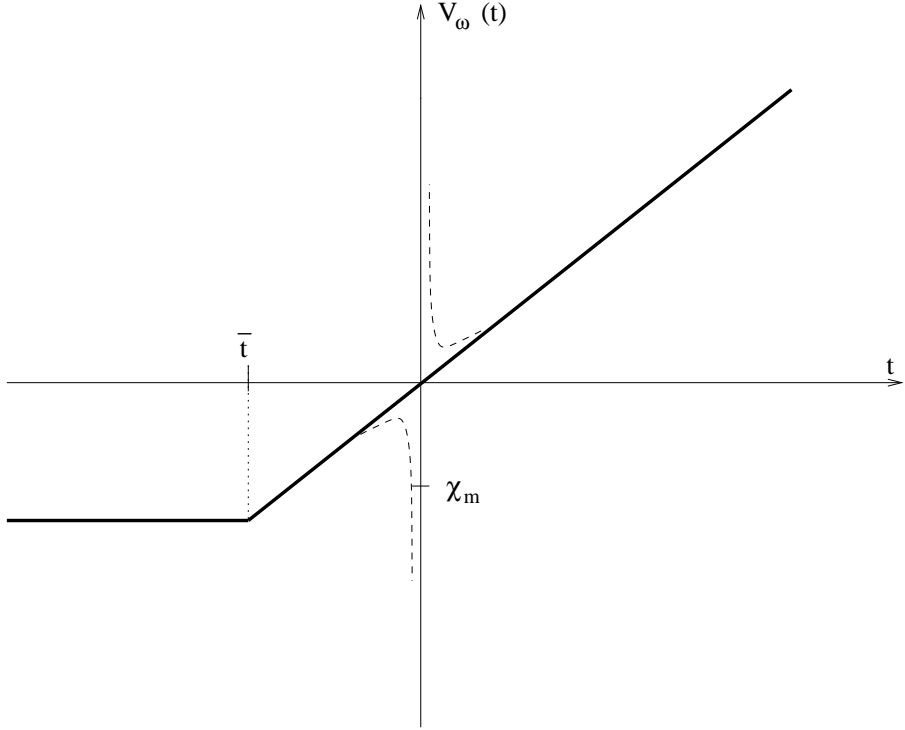,width=12cm}  
           }  
\vspace*{0.5cm}  
\caption{\it Plot of the linear potential $b\om t$ with freezing below
  $\bar{t}$. $\chi_m$ has here the interpretation of the energy in the
  Schroedinger-like problem.  Dashed lines correspond to the principle
  value regularisation of the running coupling.
\label{fig:linearpotential}}  
\end{figure}

\begin{figure}[t]  
  \vspace*{0.0cm}  
     \centerline{  
         \epsfig{figure=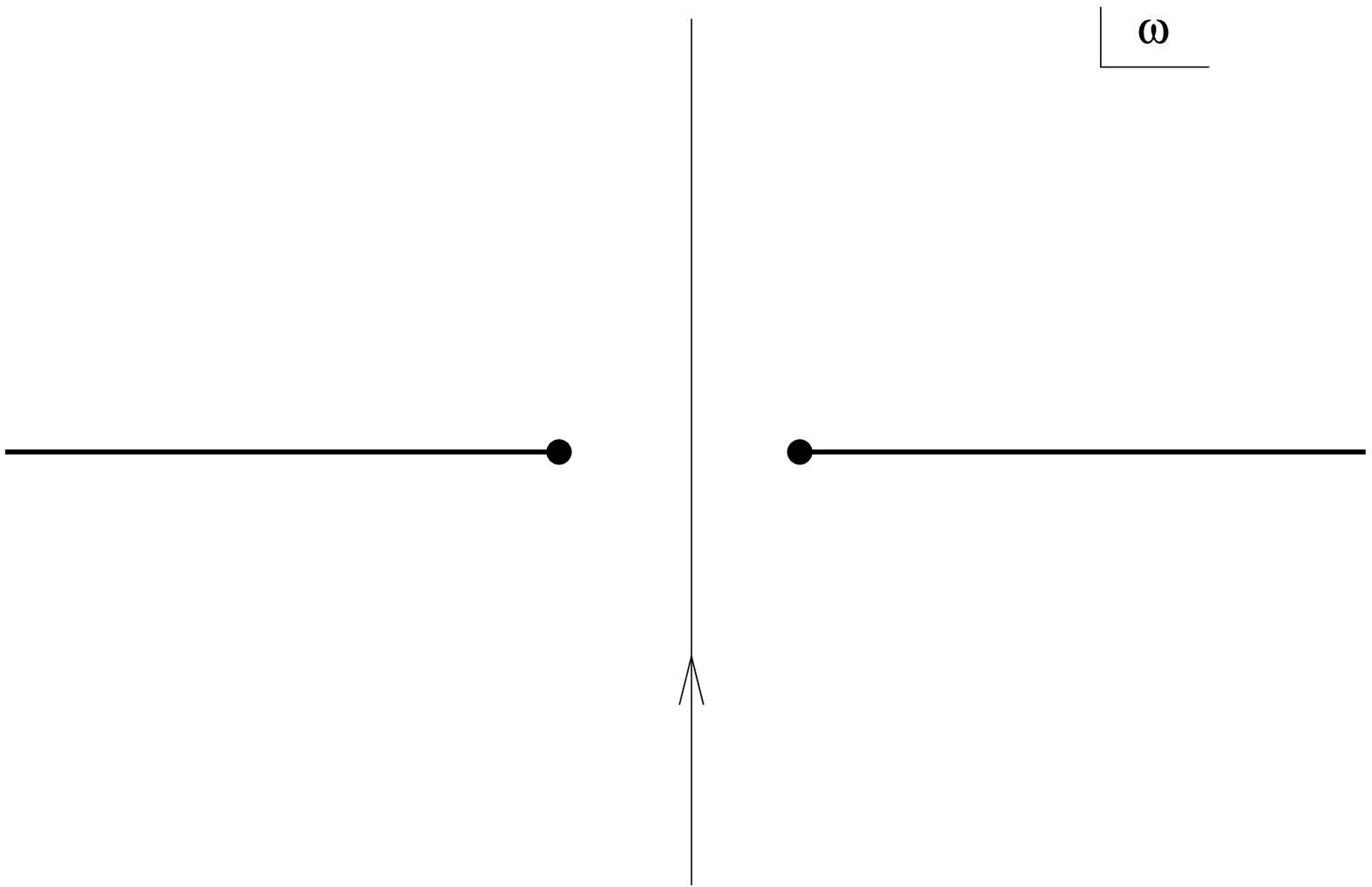,width=10cm}  
           }  
\vspace*{0.5cm}  
\caption{\it Spectrum of $\om$ values with a gap and the corresponding
  contour for the perturbative Green's function.
\label{fig:ContourOmega}}  
\end{figure}

It is now important to realise that the positive $\om$ spectrum in
(\ref{greenspectral}) is strongly dependent on the details of the
regularisation procedure around the Landau pole of the running
coupling $\asb(t)$. In fact, if one assumes that $\chi_m<0$ and
$\asb(t)$ is frozen for $t<\bar{t}$ where $\bar{t}$ is large and
negative, one realises that the spectrum has a gap for
$0<\om<{\chi_m}/(b\bar{t})$. The simplest way to see it is by using
the Schroedinger picture of Ref.~\cite{CC1}, in which the
eigenfunctions are solutions with negative energy $\chi_m$ in the
linear potential\footnote{Strictly speaking, this picture applies to
  eigenfunctions in the Airy regime of large $t$, $\frac{1}{b\om}$
  with $(b\om)^{1/3}(t-\frac{\chi_m}{b\om})$ kept fixed. That's enough
  to hint the spectral properties we need.} $\sim b\om t$ plotted in
Fig.\ref{fig:linearpotential}. We need in fact that the potential
depth $|b\om\bar{t}|>|\chi_m|$ for the continuum to begin.
Because of the gap, the Green function $\CG_{\om}(t,t_0)$ can be
continued analytically in the whole cut $\om$-plane, while the
eigenfunctions and the expression (\ref{greenprincipal}) will be only
slightly distorted for $t,t_0 > 0$ if $-\bar{t} \gg 1$. Therefore we
can define, see Fig.\ref{fig:ContourOmega}
\begin{equation}
G_{pert}(Y;t,t_0) = \int_{\epsilon-i \infty}^{\epsilon+i\infty}
\frac{d\om}{2\pi i} \CG_{\om}(t,t_0)e^{\om Y} = \left\{
  \begin{array}{rr} 
\int_{-\infty}^{0} d\om' e^{\om'Y} \CF_{\om'}(t)
\bar{\CF}_{\om'}(t_0), & (Y>0) \\
& \\
-\int_{\om_m}^{\infty} d\om' e^{\om^{\prime}Y} \CF_{\om'}(t)
\bar{\CF}_{\om'}(t_0), & (Y<0)
\end{array}
\right.
\label{gpertgap}
\end{equation}

This expression cuts off the (Pomeron-like) positive $\omega$ spectrum
completely and yields an asymptotically decreasing function for both
positive and negative $Y$'s (recall that $\chi_m<0$). Furthermore, for
$|\bar{t}| \gg 1$, one can use (\ref{greenprincipal}) and therefore
the double $\gamma$ representation (\ref{doublegammarep}) in order to
estimate it. 
 We have thus achieved our goal, that was to show that the
double $\gamma$ representation, with the $\om$-contour running over
the imaginary axis, is an acceptable definition of the perturbative
part \mbox{$G_{pert}(Y;t,t_0)$}.  In fact, while the negative $\om$
spectrum with large eigenfunctions in the large $t \gg 1 $ regime is
weakly dependent on the regularisation procedure, the positive $\om$
spectrum strongly depends on it but it is completely cut off by the
$\om$ integral being used. Since the $\om>0$ eigenfunctions are large
in the $t<0$ regime, changing the regularisation procedure to a
physical one will now
affect the perturbative part only through the tunneling from $t<0$ to
$t>0$ . Such effect is suppressed as $\sim \exp[-1/(b\asb(t))]$ for
$|t| \gg 1 $, as was the Pomeron in the estimate in Sec.2.


In other words, we propose to define the perturbative Green's function by
the $\om$-contour over the imaginary axis in Eq.~(\ref{gpertgap}) in
the double $\gamma$ 
representation with $\chi_m <0$ (followed by the analytic continuation
to $\chi_m>0$). This is an acceptable definition and cuts off the
positive $\om$-spectrum in the principal value regularisation, to be
replaced by the Pomeron one in the physical case. However there is an
intrinsic, renormalon-like ambiguity in such a definition due to the
fact that the strong coupling boundary condition still distorts the
perturbative eigenfunctions by terms which are of the relative order
$\sim \exp[-1/(b\asb(t))]$.

\subsection{ $\boldsymbol{b}$-expansion and diffusion corrections}

We have just shown that the gluon Green's function for the kernel
(\ref{eq:AsymKernel}) with a physical regularisation of $\asb(t)$, can be
written as,
\begin{equation}
G(Y;t,t_0) \; = \; G_{pert}(Y;t,t_0) \; + \;  G_{\pom}(Y;t,t_0) \; ,
\label{greendec3}
\end{equation}
where we propose to take 
\begin{equation}
\label{greenpertgam}
 G_{pert}(Y;t,t_0) \simeq \int\!\!\frac{d\om}{2\pi i} e^{\om Y} \frac{t_0}{\om}
\int^{c+i\infty }_{c-i\infty}\!\! \frac{d\gamma}{2\pi i}
\int_0^{c(\infty)}\!\! d\delta
e^{ (\gamma-1/2)t-({\gamma} + \delta -1/2) t_0 - \frac{1}{b \om}
  X(\gamma)+\frac{1}{b \om} X(\gamma+\delta)}  ,
\end{equation}
and $c(\infty)$ is a proper specification of the $\delta$ contour with
$\mathrm{Re}\frac{\delta}{\om} > 0$ .  As was mentioned in the
previous subsection this decomposition is not unique, but here we
focus on properties of the perturbative expression
(\ref{greenpertgam}) which are independent of renormalon ambiguities.

Let us now evaluate (\ref{greenpertgam}) in the regime in which
$t,t_0,1/b\om, \om Y$ are all large parameters. One way of looking
at it is to let $b\rightarrow 0$ with $\asb(t),\asb(t_0)$ and $b Y$
kept fixed, which is the $b$-expansion we are investigating. This
allows us to perform a saddle point estimate of the exponent in
(\ref{greenpertgam}) for any $\om$ value. It also ensures that the
Pomeron contribution is strongly suppressed by the tunneling
factor\footnote{In particular regimes of the $Y$ and $t$ parameters
  the present perturbative estimates are still valid for $b={\cal
    O}(1)$ but the Pomeron suppression is not strong and contaminates
  the perturbative behaviour.} (\ref{tunnelfactor}).  In such a regime
we obtain the saddle point conditions
\begin{equation}
\begin{array}{lll}
 b \bar{\om} t & = & \chi(\bar{\gamma})  \\
 b \bar{\om} t_0 & = & \chi(\bar{\gamma} + \bar{\delta})  \\ 
 b \bar{\om}^2 Y & = & X(\bar{\gamma} + \bar{\delta}) -
 X(\bar{\gamma})  \; .
\label{saddlecond}
\end{array}
\end{equation}
For instance, if we take $t_0=t$ and we use the (anti)symmetry of
$\chi(\gamma)$ ($X(\gamma)$) for $\gamma \leftrightarrow 1-\gamma$, we
can define $\bar{\gamma} = \frac{1}{2}-\bar{p}$ with the equations
\begin{eqnarray}
\bar{\delta} & = & 2 \bar{p} \nonumber \\
b \bar{\om} t & = & \chi(\frac{1}{2} -\bar{p}) \nonumber \\
b \bar{\om}^2 Y & = & - 2 X(\frac{1}{2}-\bar{p}) = 2
X(\frac{1}{2}+\bar{p})  \; ,
\label{delta}
\end{eqnarray}
and the saddle point exponent of (\ref{greenpertgam}) becomes
\begin{equation}
E(Y;t,t_0) \; = 
\bar{\om} Y - 2 \bar{p} t + \frac{2}{b\bar{\om}} X(\frac{1}{2} +
\bar{p}) \; =
\; \bar{\om} Y \; - \; \frac{2}{b\bar{\om}} \int_{\chi_m}^{b \bar{\om}
  t}\!\! p(x) dx \; ,
\label{saddleexp}
\end{equation}
where we have again used the integration by parts, analogous to
(\ref{anomrel}).  Insight into the behaviour of (\ref{saddleexp}) is
obtained by eliminating $\bar{\om}$ in Eq.~(\ref{delta}), to get
($\oms^0(t) = \chi_m/bt$ )
\begin{eqnarray}
\bar{\om} & = & \frac{\chi(\frac{1}{2}-\bar{p})}{bt} \; ,\nonumber \\
\zeta & \equiv & \frac{\chi_m Y}{b t^2} = \frac{2
  X(\frac{1}{2}+\bar{p}) \chi_m}{\chi^2(\frac{1}{2}-\bar{p})} =
g(\bar{p}) \; .
\label{dzeta}
\end{eqnarray}
\begin{figure}[t]  
  \vspace*{0.0cm}  
     \centerline{  
         \epsfig{figure=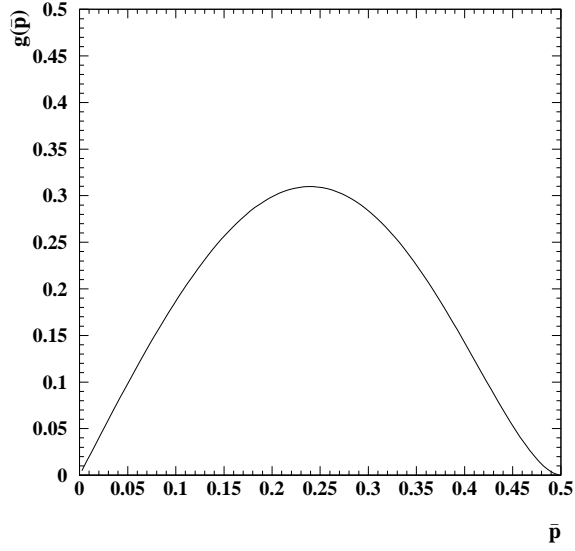,width=8cm}  
           }  
\vspace*{0.5cm}  
\caption{\it Plot of the $g(\bar{p})$ function (\ref{dzeta}) in the case 
of
  collinear model $\chi(\gamma) = \frac{1}{\gamma(1-\gamma)}$.
\label{fig:gfunction}}  
\end{figure}  
We see, Fig.~\ref{fig:gfunction},
 that for any given value of parameter $\zeta = \oms^0(t) Y/t =
b \asb \oms^0 Y$ fixed by both energy and scale, there is a value of
$\bar{p}$ which in turn determines the saddle point $\bar{\om} =
\asb(t) \chi(\frac{1}{2} - \bar{p}[\zeta])$. When $\zeta \ll 1$ is a
small parameter, the solution of (\ref{dzeta}) is driven towards the
minimum of $\chi$ , corresponding to $\bar{p}=0$. The other solution
with $\bar{p}\simeq 1/2$ ($\bar{\gamma}\simeq 0$) is instead subleading and unstable
and is thus discarded.  By expanding the
eigenvalue function $\chi$ around the minimum  we obtain,
\begin{eqnarray}
\zeta & \simeq & 2\bar{p} \frac{1+\frac{1}{3} D
  \bar{p}^2}{(1+D\bar{p}^2)^2} \simeq 2 \bar{p} (1-\frac{5}{3} D
\bar{p}^2 + \frac{7}{3} D^2 \bar{p}^4 - \frac{9}{5!}\frac{\chi_m^{(4)}}{\chi_m} \bar{p}^4 + \cdots) \nonumber \\
\bar{p} & \simeq & \frac{\zeta}{2} (1 + \frac{5}{12}  D \zeta^2 +
\frac{3}{8} D^2 \zeta^4 + \frac{9}{80}\, \frac{\chi_m^{(4)}}{4! \chi_m}\zeta^4 + \cdots) \nonumber \\
\bar{\om} & = & \oms^0(t) (1 + D \frac{\zeta^2}{4} + \frac{5}{24} D^2 \zeta^4 + \frac{1}{16} \, \frac{\chi_m^{(4)}}{4! \chi_m} \zeta^4 + \cdots) \; ,
\label{expansions}
\end{eqnarray}
and the saddle point exponent (\ref{saddleexp}) becomes
\begin{equation}
E(Y;t,t) \; = \; 2 \bigg[ \chi(\frac{1}{2}-\bar{p}) \frac{Y}{b t}-
\bar{p}t \bigg]=
\oms^0(t) Y \bigg[ 1 + \frac{1}{12} D \zeta^2 +  \frac{1}{24}D^2 \zeta^4 + \frac{1}{120} \frac{\chi_m^{(4)}}{\chi_m} \left(\frac{\zeta}{2}\right)^4  +  \cdots \bigg] \, .
\label{saddleexpsol}
\end{equation}

The result (\ref{saddleexpsol})
exhibits the existence of a power series in the parameter $z=D
\zeta^2$, previously emphasised \cite{CMT} in the context of the Airy
model. The first correction of this series yields the term $\sim
D(\oms^0 Y)^3/t^2$ found in various papers, for example see
\cite{KOVMUELLER,ABB,LEVIN}.  Further corrections, subleading 
in $Y$,
come from integrating 
the fluctuations around the saddle point  (\ref{delta}).  The
fluctuation matrix has two positive and one negative eigenvalue, roughly 
corresponding to $\gamma$ and $\om$ fluctuations along the imaginary
axis and to $\delta$ fluctuations along the real axis (see Appendix
A). As a result one obtains the following normalisation factor

\begin{eqnarray}
N(t,Y)  & = &   \sqrt{\frac{b\bar{\om}}{-4 \pi
    \chi^{\prime}(\bar{\gamma})} } \frac{1}{\sqrt{1 + 
 \frac{ \chi^{\prime}(\bar{\gamma})}{\chi_m }\zeta}} =
\frac{1}{\sqrt{4\pi D \om_0(t) Y }}
(1 + a D \zeta^2 + \cdots) \nonumber \\
& \mathrm{with} & \nonumber \\
& & a  =  \frac{5}{12} - \frac{1}{24} \frac{\chi_m
  \chi_m^{(4)}}{(\chi_m^{''})^2}
\label{normfactor}
\end{eqnarray}
which shows further dependence on $z=D\zeta^2$, of the type $\sim
Y^2$.  From higher order fluctuations one finds also further
subleading corrections with linear dependence on $Y$. In appendix B we
show how to obtain them in the specific case of the Airy diffusion
model (see Sec. 4).
\begin{figure}[t]  
  \vspace*{0.0cm}  
     \centerline{  
         \epsfig{figure=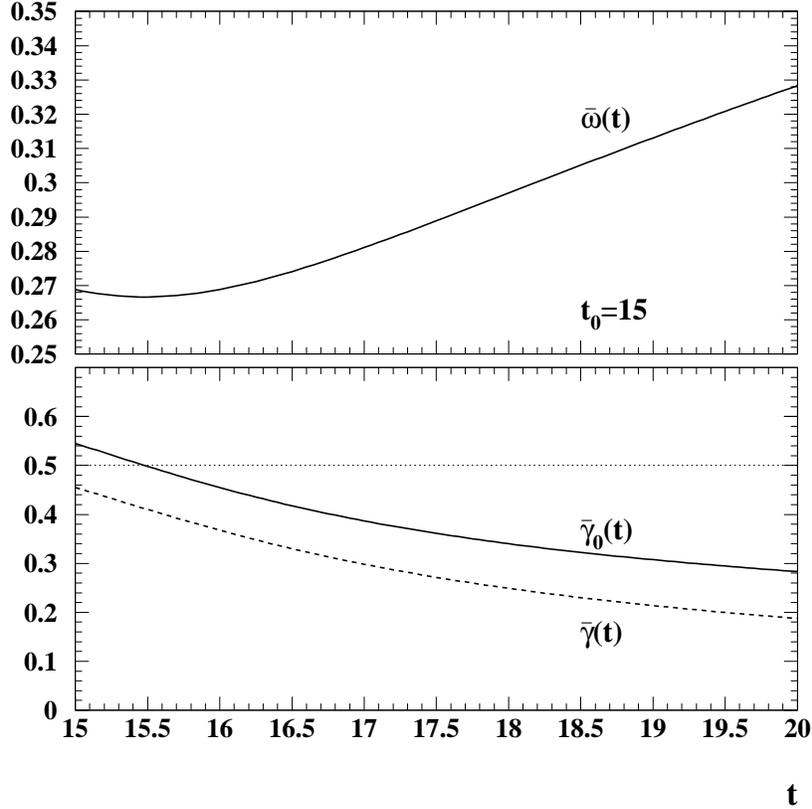,width=12cm}  
           }  
\vspace*{0.5cm}  
\caption{\it Upper plot: the saddle point solution $\omb(t)$ as a
  function of $t$ for the collinear model. Lower plot: saddle point
  solutions $\gamb(t)$ (dashed line) and $\gamb_0(t)$ (solid line) .
  The parameters are : $b=1$, rapidity $Y=5$ and scale $t_0=15$.
\label{fig:ombcoll}}  
\end{figure} 
In the BFKL case the term linear in $Y$ is calculated in \cite{CCSS3} and provides the shift
\begin{equation}
\label{eq:YShift}
\delta \omega_s^{(2)} Y = \frac{1}{12} (b\asb)^2 D \omega_s^0 Y \left[\frac{15}{4}-\frac{7}{4}\frac{\chi_m \chi_m^{(4)}}{\chi_m^{''2}}-\frac{1}{4}\frac{\chi_m^2 \chi_m^{(4)2} }{\chi_m^{''4}}+\frac{1}{8}\frac{\chi_m^2 \chi_m^{(6)}}{\chi_m^{''3}} \right] \; ,
\end{equation}
to the hard Pomeron exponent due to the diffusion corrections.

Let us note that the saddle point (\ref{dzeta}) can also be studied
for $\zeta = {\cal O}(1)$ and $0 < \bar{p} < \frac{1}{2}$. Due to the
expansion of $\chi$ around $\gamma = \frac{1}{2}$ and to the collinear
behaviour $\chi(\gamma) \simeq \frac{1}{\gamma} + \frac{1}{1-\gamma} +
\cdots$, the function $g(\bar{p})$ in the r.h.s of (\ref{dzeta}) is
nonnegative and vanishes for both $\bar{p} \rightarrow 0$ ($g(\bar{p})
\sim c \bar{p}$) and for $\bar{p} \rightarrow \frac{1}{2}$, (
$g(\bar{p}) \sim (\bar{p}-1/2)^2 \log(\bar{p}-1/2)$), see
Fig.~\ref{fig:gfunction}.  Therefore it has a maximum at some value of
$\bar{p} = p_c$. Correspondingly, for $\zeta \ge \zeta_c =
g(\bar{p}_c)$ , the saddle point(s) become complex conjugate, and the
exponent $E(\om Y,\zeta)$ acquires a branch point. In the case of BFKL
the critical value $\zeta_c\simeq 0.264$. Such behaviour was
hinted at in \cite{CMT} and is here shown to be quite general.

To summarize, we have shown that the perturbative Green's function in
the $b$-expansion takes the form
\begin{equation}
\label{eq:GreenPertGen}
G_{pert}(Y;t,t_0) \; \simeq \;
\frac{n(\zeta,t/t_0)}{2\sqrt{\pi D \oms^0(t) Y}} 
\, \exp[\oms^0(t)\, Y\, {\cal E}(\zeta,t/t_0) ] \; ,
\end{equation}
where $\zeta = \oms^0(t_0) Y/t_0$ and for $t = t_0$,
\begin{equation}
\label{eq:NormEps}
n(\zeta)  = 1 + a D \zeta^2 + \dots, \hspace{2cm} {\cal E}(\zeta)=1+\frac{1}{12} D \zeta^2 +\dots \; ,
\end{equation}
are given by power series in $\zeta$ with convergence radius $\zeta_c$
which is equal to the maximum of  the function $g(\bar{p})$, see 
Fig.~\ref{fig:gfunction}. Note that the
exponent in Eq.~(\ref{eq:GreenPertGen}) is $\order{\frac{1}{b}}$ at
fixed values of $bY$, $\asb(t)$ and $\zeta$, while the normalisation
is $\order{1}$ and the fluctuations are $\order{b^n}$, $n \ge 1$, as
is natural from the role of $b$ as a semiclassical expansion
parameter.
\begin{figure}[t] 
  \vspace*{0.0cm}  
     \centerline{  
         \epsfig{figure=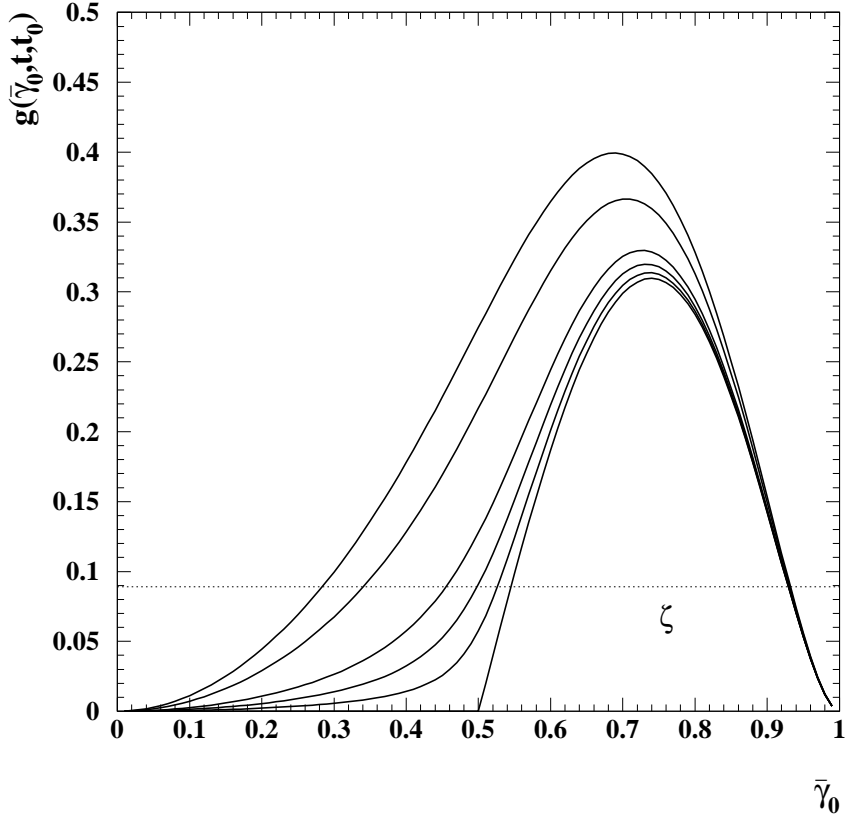,width=12cm}  
           }  
\vspace*{0.5cm}  
\caption{\it Function $g(\gamb_0;t,t_0)$ from Eq.~(\ref{geng}) plotted
  versus $\gamb_0$ for different values of $t$:
  $15,15.2,15.5,16,18,20$.  The value of the maximum increases with
  increasing $t$.  The dotted line represents the value of the
  parameter $\zeta = \chi_m Y /(b t_0^2)$. The parameters are the same
  as in Fig.~\ref{fig:ombcoll}.
\label{fig:gfuncoll}}  
\end{figure} 

The above presented analysis can be easily extended to the case of
different scales $t > t_0$, in order to reproduce the collinear limit.
The exact solution to the set of equations (\ref{saddlecond}) is
rather complicated to be found analytically for the general case of
non-equal scales; it can be however easily solved numerically.  In
Fig.~\ref{fig:ombcoll} we show the solution for the $\omb, \gamb$ and
$\gamb_0 = \gamb + \delta$ in the case of the collinear model
$\chi(\gamma) = 1/(\gamma(1-\gamma))$ as a function of scale $t$
starting from $t=t_0$. The saddle point exponent $\omb$ first
decreases with $t$, then has a minimum $\omb_m = \oms^{(0)}(t_0)$ at
certain $t_m$ and then increases. The solution for the $\gamb_0$
starts from the value $\gamb_0 > 1/2$ but then moves towards zero as
the $\gamb$ also does. The point at which $\gamb_0(t_m)=1/2$ is
exactly the minimum of the saddle point exponent $\omb(t)$.  In the
regime when $t \gg t_0 \gg 1$, both solutions tend to zero
($\gamb,\gamb_0\rightarrow 0$) and one can recover the correct asymptotic behaviour by
making the collinear approximation in (\ref{saddlecond})
\begin{eqnarray}
&  b \omb t  & \simeq \frac{1}{\gamb} \; \;  , \; \;   b \omb t_0
\simeq \frac{1}{\gamb_0}  \nonumber \\
&  b \omb^2 Y & \simeq  \ln \frac{\gamb_0}{\gamb}  \simeq \ln
\frac{t}{t_0}
\label{collinear}
\end{eqnarray}
which gives 
\begin{equation}
\label{eq:CollExp}
\omb \simeq \sqrt{\frac{1}{bY} \ln \frac{t}{t_0}} \hspace{1cm}
\mathrm{and} \hspace{1cm} 
E(Y;t,t_0) \simeq 2 \sqrt{\frac{1}{b} Y \ln \frac{t}{t_0}} \; .
\end{equation}

The function $g(\bar{p})$ is generalised in this case too, and depends
additionally on $t$ and $t_0$
\begin{equation}
g(\gamb_0;t,t_0) \equiv \chi_m \gamb_0^2 (1-\gamb_0)^2 \, \ln\left[
  \frac{\gamb_0}{1-\gamb_0} \frac{\frac{1}{2}+\sqrt{\frac{t_0}{t}}
    \sqrt{(\gamb_0-1/2)^2 + \frac{\Delta t }{4 t_0}) }}{\frac{1}{2}-
    \sqrt{\frac{t_0}{t}} \sqrt{(\gamb_0-1/2)^2 + \frac{\Delta t }{4
        t_0}) }} \right]  \; ,
\label{geng}
\end{equation}
where we have defined $\Delta t \equiv t-t_0$.  In
Fig.~\ref{fig:gfuncoll} we show it as a function of $\gamb_0$ for
different values of the scales $t$ and $t_0$. Again the value of the
parameter $\zeta =\chi_m Y/(b t_0^2) = g(\gamb_0,t,t_0) $ will
determine the solutions for the $\gamb_0$ and the related quantities.
The solution will be given by the lower value of $\gamb_0$ and it will
quickly pass $1/2$ and shift towards $0$ with increasing asymmetry
$\Delta t$. The maximum of the curves $g(\gamb_0;t,t_0)$ also
increases with $\Delta t$ which means that in the collinear situation
when $t \gg t_0 \gg 1$ one has a real solution for quite large
positive $\zeta$ and that the critical value $\zeta_c \sim \ln t/t_0$.
Note, however that due to the peculiar shape of the function
$g(\bar{\gamma}_0;t,t_0)$ (Fig.~\ref{fig:gfuncoll}) for $\zeta
\rightarrow 0^+$ both $\bar{\gamma}$ and $\bar{\gamma}_0$ drift
towards zero, as in Eq.~(\ref{collinear}), and that the exponent in
Eq.~(\ref{eq:CollExp}) implies a singular behaviour of the exponent
function ${\cal E}$ in Eq.~(\ref{eq:GreenPertGen})
\begin{equation}
\label{eq:CollEpsilon}
{\cal E}(\zeta,t/t_0) \simeq 2 \sqrt{\frac{\log t/t_0}{\chi_m \zeta}} \; , \hspace{2cm} \; \zeta = \oms(t) Y/t_0 \;  \; .
\end{equation}
Therefore, for $t>t_0$, the exponent function ${\cal E}$ is no longer
a power series in $\zeta$ around $\zeta=0$, but should be rather
expanded around some nontrivial value of $\zeta$ (like
$\zeta=g(\frac{1}{2};t,t_0)$, for which $\bar{\gamma_0} \simeq
\frac{1}{2}$), so as to stay away from both $\zeta=0$ and
$\zeta=\zeta_c$.

On the other hand, if $(t-t_0)/t_0 = b\asb(t_0) \Delta t \ll 1$ the
exponent in Eq.~(\ref{eq:CollExp}) takes up the double log DGLAP form
$\sim 2\sqrt{Y\asb(t_0) \Delta t}$ with frozen coupling (one cannot
distinguish $\asb(t)$ from $\asb(t_0)$ for such  values of
$\Delta t$). Since this expression is finite at fixed $\asb(t_0)Y$,
one can still perform an expansion of the exponent in $b$ in this
small $\Delta t$ regime, as proposed for a numerical procedure in
Sec.5.

\section{Explicit Green's function in the diffusion model}

The diffusion model with running coupling of Eq.~(\ref{airyequation})
has been discussed in the literature \cite{AIRY} in various ways,
starting from its equivalence \cite{CC1} with a Schroedinger-like
problem in $t$ space. We reconsider it here with a purpose of
providing a representation for it which makes the $b$-expansion and
the evaluation of diffusion corrections quite explicit.


We start from Eq.~(\ref{greenpertgam}), with $\chi_m<0$ and with the
specification of the $\delta$ contour provided by
Eq.~(\ref{greenprincipal}), according to which \mbox{Re$(\delta/\om)$}
should be positive. The exponent in (\ref{greenpertgam}) in the case
of the diffusion model has the following form
\begin{equation}
E := \om Y + (\gamma-1/2) \Delta t - \delta t_0 + \frac{\delta
  \chi_m}{b \om} + \frac{1}{2}\frac{\delta \chi_m^{\prime\prime}}{b
  \om} (\gamma-1/2)^2 + \frac{1}{2}\frac{\delta^2
  \chi_m^{\prime\prime}}{b\om} (\gamma-1/2) +
\frac{1}{6}\frac{\chi_m^{\prime\prime}}{b\om} \delta^3 \; .
\label{expdiff}
\end{equation}
We notice that, since $X(\gamma)$ is cubic in $\gamma$ for the
diffusion model, $X(\gamma+\delta)-X(\gamma)$ is quadratic in
$\gamma$, with coefficient of the quadratic term given by
$\frac{1}{2}\frac{\chi_m^{\prime\prime}\delta}{b\om}$, which has a
positive real part. Therefore the $\gamma$ integral, at fixed $\om$
and $\delta/\om$, converges along the imaginary axis and can be done
explicitly to yield
\begin{equation}
G_{pert}(Y;t,t_0) = \int\!\! \frac{d\om e^{\om Y}}{2 \pi i}\!
\int_{0}^{\infty} \! dx \sqrt{\frac{b t_0}{2 \pi \chi_m^{\prime
      \prime}}} \, \exp\bigg[ -x^2(\om\frac{t+t_0}{2 t_0}-\alpha_0
\chi_m) + \frac{\Omega \om^2 x^6}{4} - \frac{(\Delta t)^2 b t_0}{2
  \chi_m^{\prime\prime}x^2}\bigg]
\label{gpertxvar}
\end{equation}
where we have introduced the variables
\begin{equation}
x^2 = \frac{\delta t_0}{\om},\hspace{1cm} \Delta t = t - t_0 \, ,
\hspace{1cm}\Omega(t_0) = \frac{1}{6}\frac{\chi_m^{\prime\prime}}{b
  t_0 t_0^2} \, , \hspace{1cm} \bar{\alpha}_0 = \frac{1}{b t_0} \, .
\end{equation}
Surprisingly enough, the $\om$ dependence of the exponent in
Eq.~(\ref{gpertxvar}) is still quadratic at fixed $x^2 \sim
\delta/\om$, with positive coefficient of the quadratic term.
Therefore the $\om$ integral converges along the imaginary axis and
can be done explicitly to yield
\begin{equation}
G_{pert}(Y;t,t_0) = \int_0^{\infty} \!\!
\frac{dx}{x^3\sqrt{\pi\chi_m^{\prime\prime}}}
\sqrt{\frac{ b t_0}{2 \pi \Omega(t_0)}} \exp
\bigg[-\frac{(Y-x^2\frac{t+t_0}{2 t_0})^2}{ \Omega(t_0) x^6} +
\frac{\chi_m x^2}{ b t_0} - \frac{(\Delta t)^2 b t_0}{2
  \chi_m^{\prime\prime} x^2} \bigg] \, ,
\end{equation}
where we recall that the $x$ integral converges because we have set
$\chi_m<0$ . Finally, by introducing the new integration variable $\xi
= \frac{|Y|}{x^2}\frac{2 t_0}{t+t_0}$, we obtain $(\tilde{t} \equiv
\frac{t+t_0}{2})$

\begin{align}
G_{pert}(Y;t,t_0) &  =  \frac{t_0}{\tilde{t}} \int_0^{\infty} \!\!
\frac{d\xi}{\sqrt{2\pi \chi_m^{\prime\prime} \asb(\tilde{t}) |Y|}}
\frac{1}{\sqrt{\pi \Omega(\tilde{t}) |Y|} } \nonumber \\
& \quad \times \; \exp \bigg[-\frac{\xi (\xi \mp 1)^2}{\Omega(\tilde{t}) |Y|} +
\frac{\oms(\tilde{t}) |Y|}{\xi} - \frac{(\Delta t)^2 \xi}{2
  \chi_m^{\prime\prime}\asb(\tilde{t})  |Y|}\bigg] 
\label{gpertsolxi}
\end{align}
where $\oms(\tilde{t}) = \frac{\chi_m}{b\tilde{t}} < 0$ and the $-(+)$
sign holds according to whether $Y>0$ ($Y<0$). \\

Several features of Eq.~(\ref{gpertsolxi}) are worth noting. First,
the solution decreases for $|Y| \rightarrow \infty$ in both
directions, as expected from Eq.~(\ref{gpertgap}), but more strongly
for $Y\rightarrow -\infty$. Secondly, it satisfies the following
boundary condition
\begin{equation}
G(Y=0^+;t,t_0) - G(Y=0^-;t,t_0) = \delta(t-t_0) \; ,
\label{gboundcon}
\end{equation}
but the $Y=0^-$ contribution is non-vanishing. As a matter of fact we
have
\begin{eqnarray}
G(Y=0^+;t,t_0) & = & \int_0^{\infty}\! d\xi\,
\delta(\sqrt{\xi}(t-t_0)) \, \delta(\sqrt{\xi}(\xi-1)) \, = \, 2
\delta(t-t_0) \nonumber \\ 
G(Y=0^-;t,t_0) & = & \int_0^{\infty}\! d\xi\,
\delta(\sqrt{\xi}(t-t_0)) \, \delta(\sqrt{\xi}(\xi+1)) \, = \,
\delta(t-t_0)  \; ,
\label{gboundconsep}
\end{eqnarray}
and this means, according to Eq.~(\ref{gpertgap}), that the projector
over the negative (positive) spectrum is 2(-1) in this case. Third,
apart from the overall $t_0/\tilde{t}$ factor(due to the asymmetrical
role of the running coupling in the kernel), the $t$ and $t_0$
dependence occurs mostly through the parameter $\tilde{t} =
\frac{t+t_0}{2}$ corresponding to the scale $k k_0$ for the hard
process.  Finally, the $b$-dependence is nicely summarized by
introducing the parameter
\begin{equation}
z \; = \; b^2 \asb(\tilde{t})^2 \frac{\chi_m \chi_m^{\prime\prime}}{2}
(\asb(\tilde{t}) Y)^2 \; = \; 3 \Omega(\tilde{t}) \oms Y^2 \, ,
\label{zvar}
\end{equation} 
in terms of which Eq.~(\ref{gpertsolxi}) reads
\begin{equation}
G_{pert}(Y;t,t_0) = \frac{t_0}{\tilde{t}} \int_0^{\infty} \!
\frac{d\xi}{2\pi \sqrt{D \, |z|/3}}
\exp \bigg\{\oms(\tilde{t}) |Y| \bigg[\frac{\xi(\xi\mp 1)^2}{(-z)/3} +
\bigg(\frac{\Delta t}{2\tilde{t}}\bigg)^2 \frac{\xi}{(-z)} +
\frac{1}{\xi}\bigg]  \bigg\}
\label{gpertsolz}
\end{equation}
for $Y>0$ ($Y<0$) respectively. This equation provides the almost explicit
representation of $G$ we were looking for.

\begin{figure}[t]  
  \vspace*{0.0cm}  
     \centerline{  
         \epsfig{figure=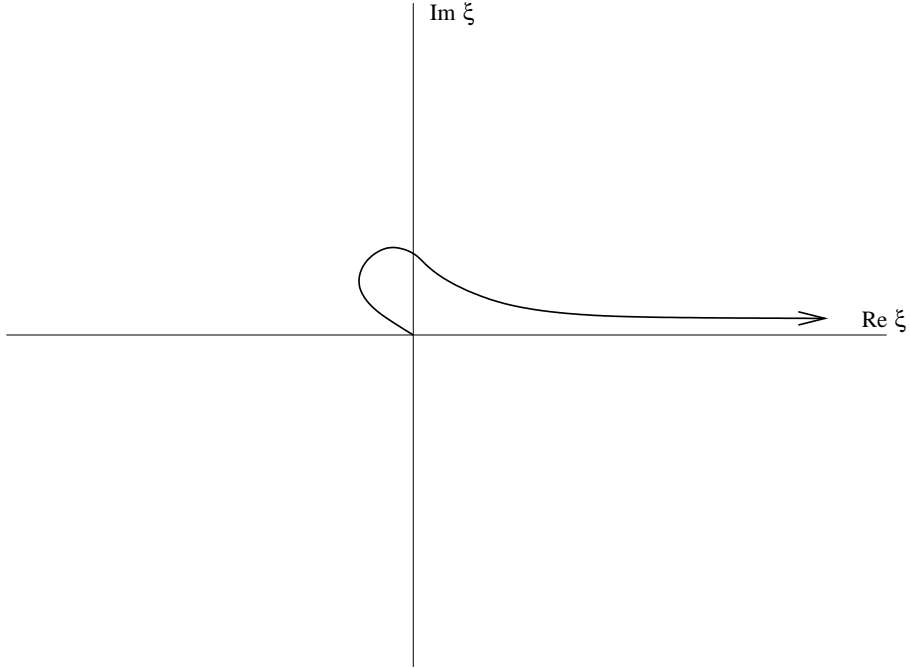,width=12cm}  
           }  
\vspace*{0.5cm}  
\caption{\it A possible continuation to $\chi_m>0$ of the contour  in
  the $\xi$ plane for the integral (\ref{gpertsolz}) representing the
  perturbative solution of the Green's function in the diffusion
  model.
\label{fig:xicontours}}  
\end{figure}

Before analyzing Eq.~(\ref{gpertsolz}) in more detail, let us consider
the analytic continuation of Eq.~(\ref{gpertsolxi}) to the positive
values of $\chi_m$. Only the $\xi=0$ region of the integral is
affected. In this case the contour in the $\xi$-complex plane has to
be distorted so as to reach the end point $\xi=0$ from the
\mbox{Re\,$\xi<0$} region, and this can be done in several ways, see
Fig.\ref{fig:xicontours}.  Since the measure of the small $\xi$ region
vanishes, this contribution is not expected to introduce sizeable
ambiguities, which are roughly given by solutions of the homogeneous 
equation with $\om=0$.

The large positive $Y$ behaviour of Eq.~(\ref{gpertsolz}) at fixed
value of $z$ is determined by the saddle point
\begin{equation}
f(\xi,\frac{\Delta t}{t}) \; = \; \xi^2 \bigg[(3\xi-1)3(\xi-1) +
\bigg(\frac{\Delta t}{2 \tilde{t}} \bigg)^2  \bigg] = -z \; .
\label{zfun}
\end{equation}
For $z<0$ ($\chi_m<0$) there are four saddle points, the leading one
being close to $\xi=1$. For $z>0$ ($\chi_m>0$) the saddle points close
to $\xi=0$ (contributing to the completeness relation) become complex
and two real ones remain. The leading one at
$\bar{\xi}\simeq1-\epsilon$ for $\Delta t \ll \tilde{t}$ and $z \ll 1$
is equal to
\begin{equation}
\bar{\xi} \; \simeq \; 1 - \frac{1}{6} \bigg(z(\tilde{t}) +
\frac{7}{12}z(\tilde{t})^2   +
\bigg(\frac{\Delta t}{2\tilde{t}} \bigg)^2 \bigg) \; ,
\label{leadsaddle}
\end{equation}
and yields the exponent
\begin{equation}
E(Y;t,t_0) \; = \; \oms(\tilde{t}) Y \bigg[1 + \frac{z(\tilde{t})}{12}
+ \frac{z(\tilde{t})^2}{24}
- \bigg(\frac{\Delta t}{2 \tilde{t}} \bigg)^2
\bigg(\frac{1}{z(\tilde{t})} + \frac{1}{6}\bigg) + \dots \bigg] \; ,
\label{exponent}
\end{equation}
thus confirming the results of Ref.~\cite{CMT} and generalizing them to
arbitrary scale dependence.
\begin{figure}[t]  
  \vspace*{0.0cm}  
     \centerline{  
         \epsfig{figure=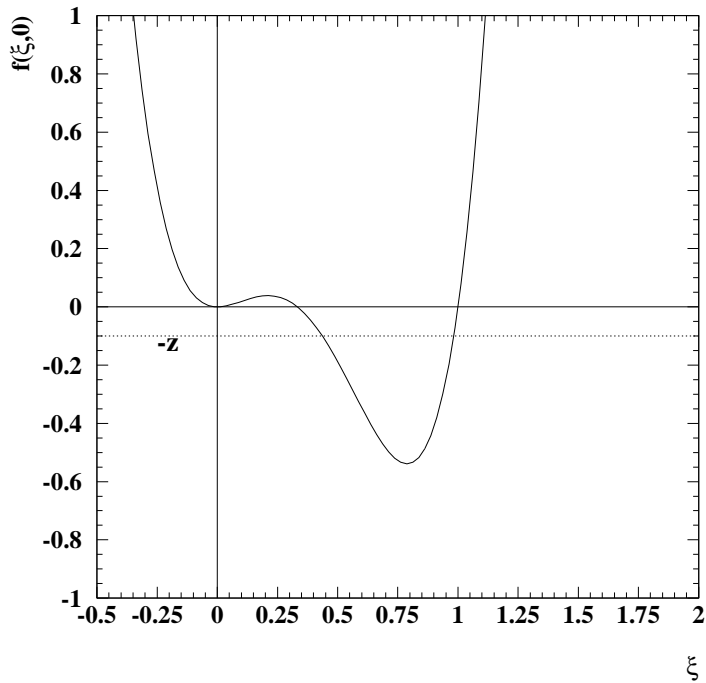,width=10cm}  
           }  
\vspace*{0.5cm}  
\caption{\it Plot of the ``saddle point'' function $f(\xi,\Delta t /
  \tilde{t} = 0)$ versus $\xi$, (\ref{zfun}). The dotted line
  represents the value of $-z$ variable, (\ref{zvar}). The
  intersection points define  the saddle point solutions for the
  exponent in solution (\ref{gpertsolz}).
\label{fig:fxifunction}}  
\end{figure}  

By taking into account the fluctuations around the saddle point we can
also obtain the $\sim Y^2$ corrections (apart from the already known
$\sim Y^3$ ones of Eq.~(\ref{exponent})). It turns out that by
considering the third order derivative of the exponent in
(\ref{gpertsolz}) one can also identify the correction linear in $\sim
Y$ which corresponds to the second order shift of $\oms(t)$.  The
details of the calculation are provided in appendix B and yield the
result $\delta \oms^{(2)} Y = \frac{5}{32} (b\asb)^2 \asb \chi_m^{''}
Y$, which reproduces the first term of (\ref{eq:YShift}).

The importance of the parameter $z(\tilde{t})$ is confirmed by the
above procedure.  In fact, above some critical value $z_c(\Delta t
/\tilde{t})$ the saddle points collide and become complex, implying a
singularity of the saddle point evaluation and thus a change of regime
occurring for $\oms(t) Y \simeq \frac{\tilde{t}}{\sqrt{D}}\sqrt{z_c}$.
 The critical value of
$z_c$ is $\frac{\Delta t }{\tilde{t}}$-dependent and ranges from
$\frac{1}{2}(\frac{1}{2}+\frac{\sqrt{3}}{3})$ for $\Delta t =0 $ to
$z_c=0$ for $\frac{\Delta t }{2 \tilde{t}}=1$, ($t_0 \ll t$).  This
has to be contrasted with the results for the case of collinear model
(see subsection 3.2 and discussion about Eq.~(\ref{geng})) where the
critical value of $z_c$ is always positive and has the asymptotic
behaviour $z_c \sim \ln^2 t/t_0$ for $t \gg t_0 \gg 1 $.  The fact
that $z_c=0$ for the Airy model in the collinear regime just means
that this approximation is no longer valid.  In fact the validity of
Airy model is strongly limited to  small values of $\Delta t$ .
Finally, for $z \gg 1 $, the complex saddle points at $\xi = \pm
\frac{1}{\sqrt{3}} (e^{\pm i\pi} \, z)^{1/4}$ dominate the exponents,
which become in that case
\begin{equation}
E_{\pm}(Y;t,t_0) \; = \; \frac{4}{\sqrt{3}} \oms(\tilde{t}) Y
z^{-\frac{1}{4}} e^{\mp\frac{i\pi}{4}} = \frac{4}{\sqrt{3}}
\frac{\sqrt{Y \chi_m /b}}{D^{1/4}} e^{\mp\frac{i\pi}{4}} \; ,
\label{explargez}
\end{equation}
thus implying an unphysical oscillatory behaviour for $\oms(t)Y \gg
\tilde{t}$, as already discussed in \cite{CMT}.

\section{Numerical results}
\label{sec:num}

In addition to the analytical studies carried out above, it is also
possible to examine the $b$-expansion from a numerical point of view.
There are two main purposes to this. One is to examine the structure
of the $b$-expansion somewhat more generally, be it at higher order,
or in  the context of more general kernels.

The other aim is to establish a way of numerically defining `purely
perturbative' predictions for high-energy scattering, as well as the
potential domain of validity of these predictions. While for
leading-order BFKL calculations this is not strictly necessary given
the analytical tools at our disposition, when including higher-order
corrections, numerical methods may represent a more practical
approach.

\subsection{Numerical $\boldsymbol{b}$-expansion}

The BFKL equation (\ref{bfklll})
 for the gluon Green's function can be written in
rapidity and transverse momentum space as
\begin{equation}
\label{eq:Bfkl}
  \frac{dG(Y;t,t_0)}{dY} = \int \frac{d^2 \vec q}{\pi q^2} \,\asb(X)
  \left[\frac{k}{k'} G(Y; \ln {k'}^2, t_0) - 
    G(Y; t, t_0) \Theta(k-q) \right]
    ,\; 
  \begin{array}{l}
    k^2 = e^{t}\,,\\
    k'  = |\vec k + \vec q|\,,
  \end{array}
\end{equation}
where the scale $X$ of the coupling may be $t$ (as has been the case
so far in this article) or $\ln q^2$, or some other combination of
scales in the problem. The gluon Green's function is defined
with the following initial condition:
\begin{equation}
  \label{eq:deltafn}
  G(0;t,t_0) = \delta(t - t_0)\,.
\end{equation}
To determine the $b$-expansion of $G$, we carry out several numerical
evolutions of the initial condition, each time with a different
(small) value of $b$. For arbitrary $b$, the coupling is defined such
that $\asb(t_0)$ is independent of $b$:
\begin{equation}
\label{eq:AlphaB}
  \asb^b(t) = \frac{\asb(t_0)}{1 + (t-t_0) b \asb(t_0)}\,.
\end{equation}
Typically we use $b$ values of $-n\delta b, -(n-1)\delta b, \ldots,
n\delta b$ with $\delta b$ of order $0.01$ and $n=3$.  The smallness
of $\delta b$ ensures the absence of non-perturbative tunneling
contributions, since they are suppressed by a factor $\exp[
-\frac{1}{|b|\asb(t_0)} g(\asb)]$.  The use of negative $b$'s may seem
surprising, but actually Eq.~(\ref{eq:AlphaB}) is valid for any sign of
$b$, and the $\gamma$-representation also, provided one replaces
$t_0$ by $\asb^{-1}(t_0)/b$.  One then considers the $b$-expansion for
$\ln G(Y; t, t_0)$,
\begin{equation}
  \label{eq:Gexpn}
  \ln G(Y; t, t_0) = \sum_{i=0} b^i \ell_i(Y;t-t_0,\asb(t_0))\,.
\end{equation}
where we assume that $t-t_0 \ll t_0$.  Expression (\ref{eq:Gexpn})
corresponds to the series in Eq.~(\ref{eq:NormEps}) for the
normalisation and exponent functions in terms of the parameter
$\zeta=b\asb(t_0) \oms(t_0)Y$.  We note that this numerical
$b$-expansion differs slightly from the analytical $b$-expansion
discussed earlier. In particular while previously the expansion was
performed with $\as(t_0)$, $\as(t)$ and $b Y$ fixed, here it is
$\as(t_0)$, $t-t_0$ and $\as(t_0) Y$ that are kept fixed.

In the approximation that for the small $b$-values being considered,
$\ln G$ is well represented by a truncation of the series
\eqref{eq:Gexpn} at term $i=2n$, it is straightforward to determine
the $\ell_i$.  Thus with $n=3$ we are able to determine terms in the
expansion up to $\ell_6$.  Formally the error on the determination of
the $\ell_i$ from neglected terms is of order $(n\delta b)^{2n+1-i}
\ell_{2n+1}$, though it can be larger if there is significant
cancellation between terms.\footnote{In practice numerical rounding
  errors are also important and contribute at the level of $\epsilon /
  \delta b^i$, where $\epsilon$ is the relative machine accuracy. The
  trade-off between these two sources of error determines the optimal
  value of $\delta b$.}

\begin{figure}[htbp]
  \begin{center}
    \epsfig{file=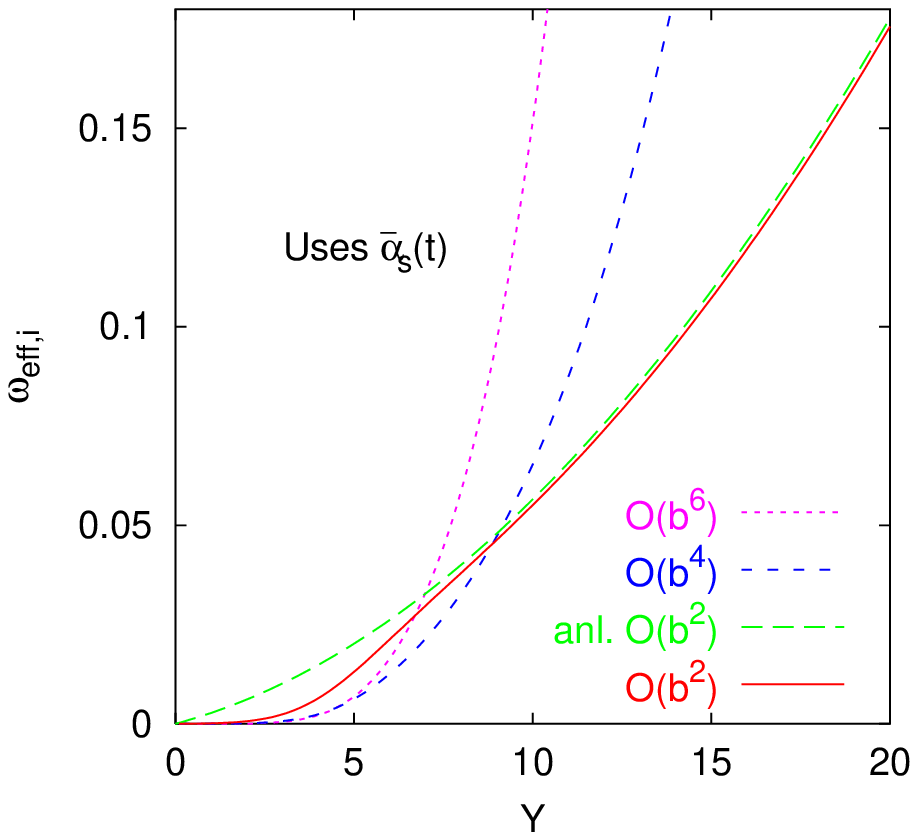,height=0.43\textwidth}
    \epsfig{file=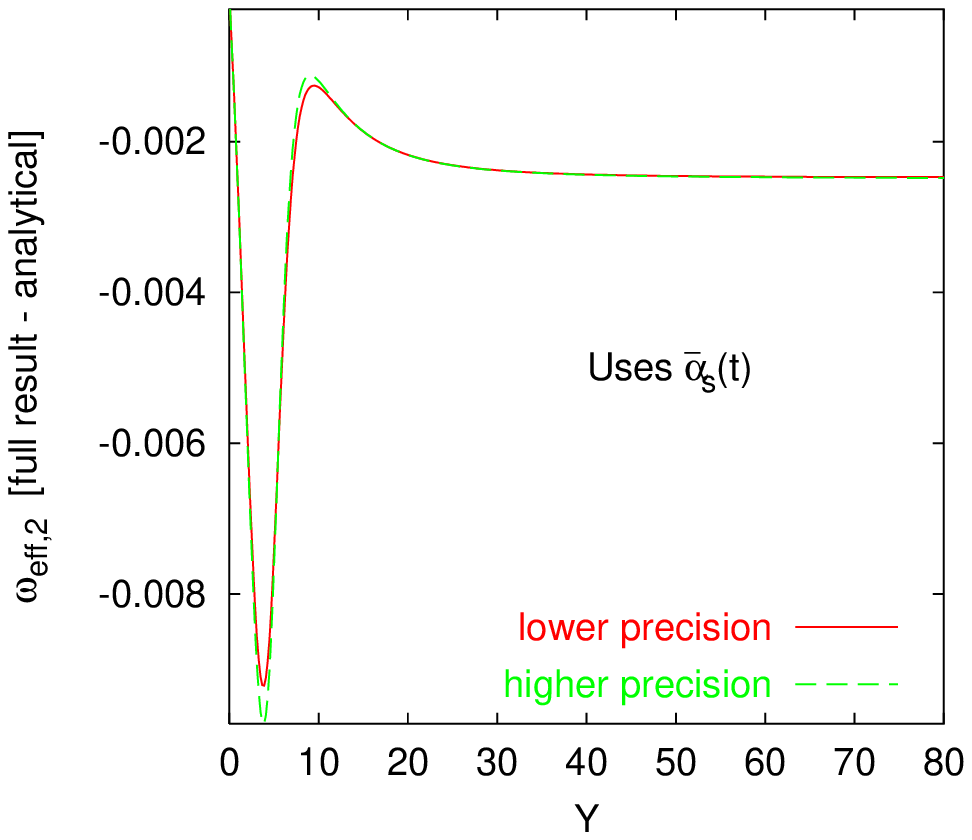,height=0.43\textwidth}
    \caption{Left: terms in the $b$-expansion of $\om_\mathrm{eff}(Y;
      t_0, t_0)$ up to order $b^6$, and additionally the asymptotic
      analytical prediction for the $b^2$ term ($b^2\as^5 Y^2$ and
      $b^2\as^4 Y$ contributions). Right: the difference between the
      full numerical result for the $b^2$ term and the asymptotic analytical
      result at lower and higher numerical precisions. In both cases
      $\asb(t_0) = 0.1$.}
    \label{fig:OmegaAst}
  \end{center}
\end{figure}

In a number of the figures that follow we will actually consider the
effective exponent
\begin{equation}
  \omega_\mathrm{eff}(Y; t_0, t_0) \equiv \frac{d}{dY} \ln G(Y; t_0, t_0)
\end{equation}
rather than the Green's function itself --- the use of $\omega_\eff$
facilitates the identification of the $Y$-dependence of certain of the
$\ell_i$, and $\omega_\eff$ can itself also be expanded in powers of
$b$,
\begin{equation}
\label{eq:Omegabexp}
  \omega_\eff = \sum_{i=0} b^i \om_{\eff,i}\,.
\end{equation}
Figure~\ref{fig:OmegaAst} shows the $\om_{\eff,i}$ for evolution with
scale $t$ and $\asb(t_0) = 0.1$. This low value of $\as$ has been
chosen so as to have a value of $\om_s(t_0) \simeq 0.277$, close to
that expected for the more physical $\asb(t_0)\simeq0.2$ once one
takes into account NLL corrections. Odd powers of $b$ have zero
coefficient, due to the $b \leftrightarrow -b$ symmetry of
Eq.~(\ref{eq:GreenPertGen}), which in turn is due to the $\gamma
\leftrightarrow 1-\gamma$ symmetry of the BFKL eigenvalue function and
eigenfunctions, Eq.~(\ref{eigenfungam}).

The $\order{b^2}$ term in the left-hand plot illustrates the
characteristic $Y^2$ dependence expected from Eq.~\eqref{saddleexpsol}
(due to the derivative the contribution $b^2 \as^5 Y^3$ to $\ln G$
gives a contribution proportional to $Y^2$ in $\om_{\eff,2}$). It is
compared to the sum of the analytically determined $b^2 \as^5 Y^2$ and
$b^2 \as^4 Y$ terms and asymptotically one sees a good agreement. The
right-hand plot shows the difference between the full numerical and
partial analytical determination of $\om_{\eff,2}$ and one sees that it
is consistent asymptotically with a constant term, which equals to
$\delta \omega_s^{(2)}/b^2  \simeq -0.002491$ from
Eq.~(\ref{eq:YShift}).
 The two curves in the right-hand plot have been determined with numerical
evolutions of different accuracies so as to illustrate the numerical
stability of the procedure. The differences visible at smaller
$Y$ values are a consequence of sensitivity to the different
approximations (widths) of the initial delta-function in $t$.

The left-hand plot also shows the $\om_{\eff,4}$ and $\om_{\eff,6}$
terms. While at low $Y$ they are suppressed, they grow much faster
with $Y$ (as $Y^4$ and $Y^6$ respectively) and quickly dominate over
the $\om_{\eff,2}$ term.
Of course, they are suppressed in Eq.~(\ref{eq:Omegabexp}) by a power
of $b$ also. But if we take $b=\order{1}$, then the $\zeta$ parameter
in Eq.~(\ref{dzeta}) is sizeable, and the importance of higher order terms
increases with $Y$, meaning that the series (\ref{eq:Omegabexp}) is
slowly converging, because of nonanalyticity at $\zeta=\zeta_c$. This
point will be discussed in more detail below and determines the
perturbatively accessible range of $Y$ for BFKL predictions.

\begin{figure}[htbp]
  \begin{center}
    \epsfig{file=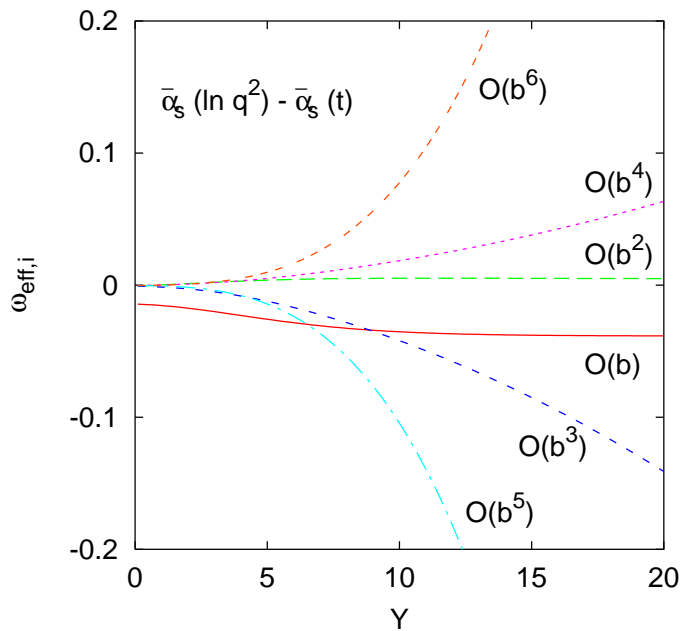,height=0.53\textwidth}
    \caption{The difference between $\om_{\eff,i}(Y; t_0, t_0)$ for
      the case of a coupling that runs as $\asb(\ln q^2)$ and one that
      runs as $\asb(t)$. Shown for $\asb(t_0) = 0.1$.}
    \label{fig:OmegaAstq}
  \end{center}
\end{figure}

It is also of interest to examine what happens to the evolution when
$\as$ is evaluated not at scale $t$ but at $\ln q^2$, which is
favoured by the NLL corrections to the BFKL equation \cite{NLLBFKL}.
The differences in the $\om_{\eff,i}$ between these two options are
shown in figure~\ref{fig:OmegaAstq}. This figure has been generated
with a Gaussian ($\propto e^{-(t-t_0)^2/2}$) initial condition as
opposed to a delta-function, because with a delta-function, at small
$Y$ one finds spuriously large contributions enhanced by logarithms of
the `width' of the delta-function.

With $\as(\ln q^2)$ one loses the $b \leftrightarrow -b$ symmetry of
the evolution and both odd and even powers of $b$ are present.
At orders $b$ and $b^2$ the effect of changing from $\as(t)$ to
$\as(\ln q^2)$ is simply to modify $\om_\eff$ by a constant ---
essentially the dynamics which led to the asymptotic $Y$-dependence in
figure~\ref{fig:OmegaAst} is independent of what scale one chooses for
$\as$.  The constant shift is trivially $\Delta = - \oms^0\frac{b
  \oms^{0}}{2}$ at order $b$, and contains both scale changing and
diffusion effects at order $b^2$.  It is only at order $b^3$ and
beyond that $Y$-dependence starts to appear, and one sees a form of
mixing between the shift in $\om_{\eff}$ which appears at relative
order $b \oms^0$ due to the scale $\as(\ln q^2)$ and the dynamical
effects which arise for any choice of scale at relative order $(b\asb
Y)^2$. This is reflected in the fact that the order $b^3$ and $b^4$
terms have a leading $Y^2$ dependence, while the $b^5$ and $b^6$ terms
have a leading $Y^4$ dependence.

\subsection{Pure perturbative predictions}

One application of the $b$-expansion is that of extracting `purely
perturbative' numerical predictions. 

One of the most common ways of obtaining BFKL predictions including
running coupling is to solve Eq.~\eqref{eq:Bfkl} (or its analogues
with various forms of higher-order correction) with a regularised
running coupling, see for example \cite{Kwiecinski:1999hv,LUND} . A
measure of the perturbative uncertainty on the prediction can then be
made by comparing different regularisations and seeing how they affect
the Green's function.

\begin{figure}[htbp]
  \begin{center}
    \epsfig{file=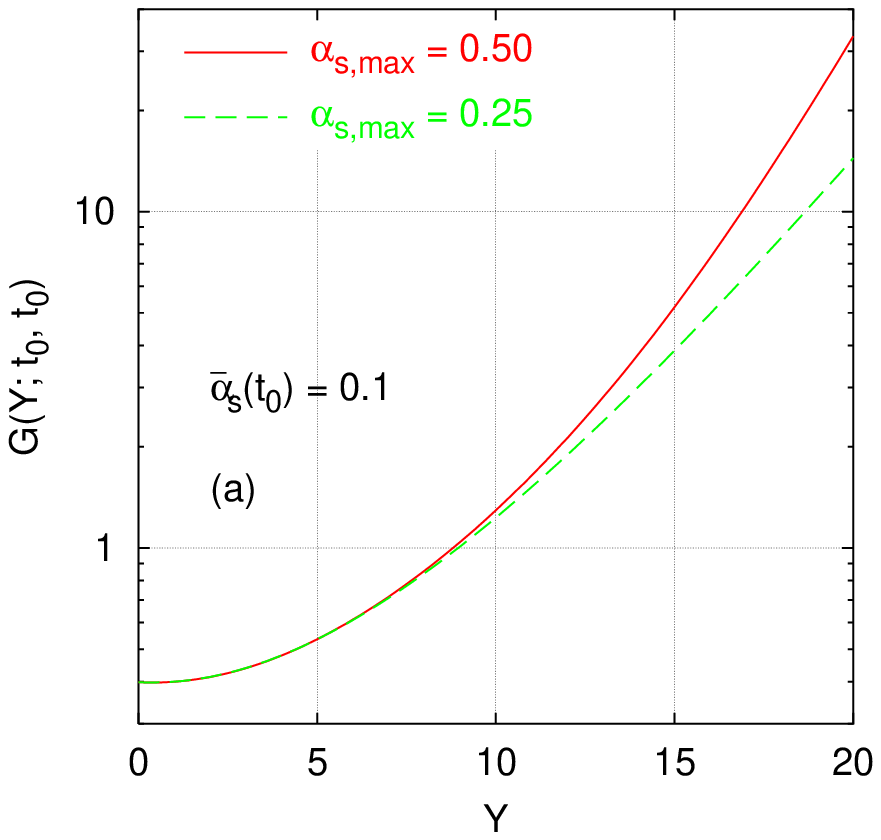,width=0.46\textwidth}\quad
    \epsfig{file=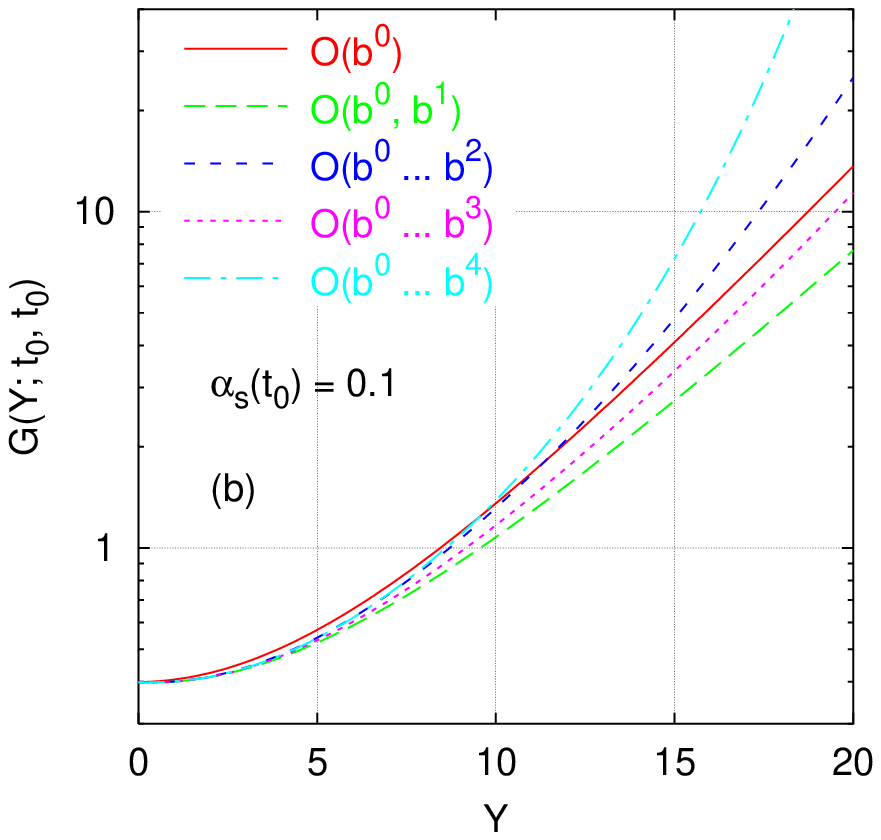,width=0.46\textwidth}
    \caption{(a) the BFKL Green's function evaluated with two
      regularisation of $\as$, a cutoff when $\asb=0.50$ and a cutoff
      when $\asb=0.25$; (b) the Green's function evaluated by
      truncating the sum Eq.~\eqref{eq:Gexpn} at various orders.}
    \label{fig:grnfns}
  \end{center}
\end{figure}

This is illustrated in Fig.~\ref{fig:grnfns}a, which shows the Green's
function evaluated by direct numerical solution of Eq.~\eqref{eq:Bfkl}
with two different regularisations of the coupling: in one the
coupling is set to zero for $t<\tbar$ where $\asb(\tbar)=0.5$, while
in the other $\tbar$ is defined by $\asb(\tbar)=0.25$. In this and
other plots in this section, we use $\as(\ln q^2)$ and the initial
condition is a Gaussian rather than a delta-function so as to have
sensible behaviour for $G$ at small $Y$.

At small and moderate $Y$ there is good agreement between the two
curves, but beyond a certain point they no longer coincide, and one
may deem this to be the limit of the perturbative prediction. However
we could quite easily have chosen a different set of regularisations
to compare to (e.g.\ with a coupling that freezes below $\tbar$) and
one would have come to a different conclusion about the point where
non-perturbative effects become important; it is difficult to make a
strong case for one regularisation scheme as opposed to another.

A second point is that for small values of $\as(t_0)$, in the
numerical calculation the transition to the non-perturbative regime
comes about by tunneling \cite{CCS1,CCSS2}, which places a limit
on the maximum accessible rapidity, $Y_\mathrm{max}$, that scales as
$1/\as(t_0)$. However there are arguments that suggest that in the
real world (as opposed to a numerical solution of a linear BFKL
equation) effects such as unitarity \cite{Balitsky2,GBMS}, or the fact
that the Pomeron is soft, will eliminate or suppress tunneling. In
such a situation the true non-perturbative limit on the maximum
rapidity is believed to scale as $1/\as^2(t_0)$ \cite{CMT}.  There is
however a problem of how practically to calculate a Green's function
beyond the tunneling point.

The $b$-expansion offers a solution to both these problems. The
question of how to determine the rapidity where perturbation theory
breaks down reduces to that of establishing when the series expansion
in $b$ stops converging. This is illustrated in
Fig.~\ref{fig:grnfns}b. One eliminates in this way the need for a
subjective decision about a regularisation of the coupling. It is
still necessary to decide where to truncate the series, but
one has an analytical understanding of the properties of the series
and the mathematical tools in such a case are well-established.

The problem of tunneling is also eliminated, because a truncated
perturbative expansion in powers of $b$ will not reproduce a
non-perturbative $e^{1/b\as}$ factor.\footnote{One may wonder whether
  tunneling might manifest itself in the series expansion through some
  form of renormalon behaviour --- it is difficult to answer this
  question properly without going to inaccessibly high orders in $b$.
  As is shown below however, in practice this issue does not seem to
  arise.} %
The maximum perturbative rapidity in the $b$-expansion is therefore
expected to behave precisely as discussed in \cite{CMT}, namely to
scale as $1/\as^2(t_0)$.

These ideas can be tested by considering a series of evolutions
for different $\as(t_0)$ values. In each case one establishes a
maximum accessible rapidity, $Y_\mathrm{max}$, in two ways:
\begin{enumerate}
\item by examining when two non-perturbative regularisations lead to
  answers which differ by more than a certain threshold;
\item by examining when the difference between different truncations
  of the $b$-expansion differ by the same threshold.
\end{enumerate}
In the first case we take the two regularisations used for
Fig.~\ref{fig:grnfns}a, while in the second case we take truncations
at order $b^3$ and $b^4$. We define the threshold as being when $|\ln
G_a/G_b| = 0.2$, where $G_{a,b}$ are the Green's function for the
different regularisations or truncations.

\begin{figure}[htbp]
  \begin{center}
    \epsfig{file=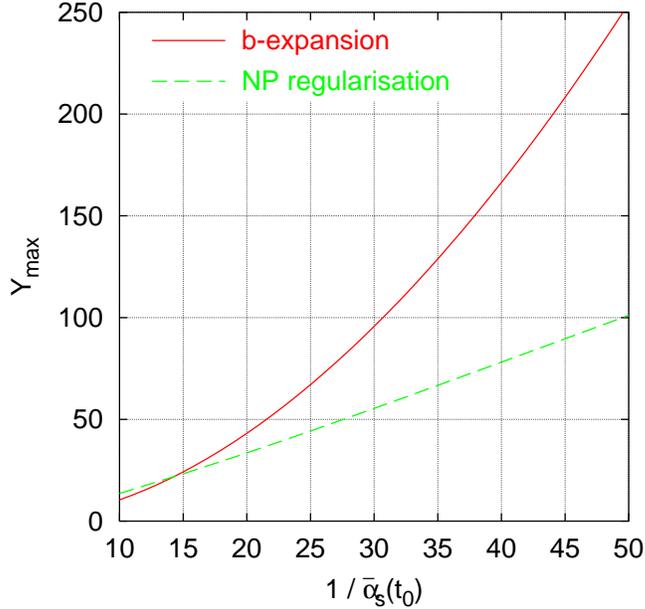,width=0.54\textwidth}
    \caption{The maximum perturbatively accessible value of $Y$, as a
      function of $1/\asb(t_0)$, determined by comparing different
      non-perturbative regularisations of $\as$, or different
      truncations of the $b$-expansion. }
    \label{fig:ymax}
  \end{center}
\end{figure}

The results are shown in Fig.~\ref{fig:ymax}. There is a
clear linear dependence on $\asb^{-1}(t_0)$ for the case of the
non-perturbative regularisation, a clear sign of tunneling being the
relevant mechanism. In the $b$-expansion $Y_\mathrm{max}$ rises
much more rapidly, in a manner quite compatible with a proportionality
to $\asb^{-2}(t_0)$. We note that the value of  coefficient in front of
$\asb^{-2}(t_0)$, of the order of $0.1$, is in perfect agreement 
with the prediction based on the critical value $\zeta_c=0.26413$
found in Sec.~3.2, which yields $Y=\frac{\zeta_c}{b\chi_m}\asb^{-2}(t_0) 
\simeq 0.104
\; \asb^{-2}(t_0) $. 

Some comments are in order. For the case of nearly leading-log BFKL
evolution that has been studied here, one could equally well have
performed a normal expansion in powers of $\as$ for fixed $\as Y$
apart from the need of the Pomeron suppression.
Indeed, it turns out that (for $t=t_0$) the $b^i \ell_i$ are functions
only of $\as Y$ and $b \as$, and one can therefore rewrite the
expansion for $\ln G$ as
\begin{equation}
  \ln G \equiv \sum_{i=0} (b \asb)^i \lambda_i(\as Y)
\end{equation}
with $\lambda_i(\as Y) = \ell_i/\as^i = (\as Y)^{i+1} c_i + (\as
Y)^{i} d_i + \dots, \hspace{0.3cm} (i \ge 1)$.  From the point of view of
Eq.~(\ref{eq:GreenPertGen}) the highest $Y$ powers in $\lambda_i \sim
(\as Y)^{i+1}$ correspond to the exponent function, the next-highest
ones ($\sim (\as Y)^{i}$) to the normalisation, and lower ones to
higher order fluctuations. Therefore, the $b$-expansion is needed
to suppress the Pomeron, but is actually in one-to-one correspondence
with an expansion in $\as$ at fixed $\as Y$.

For next-to-leading-logarithmic (and N$^n$LL) evolution the situation
is different. As is well known, the series convergence in $\as$ is
extremely poor, because of several problems stemming essentially from
large collinearly enhanced terms. To carry out an expansion in $\as$
for fixed $\as Y$ is therefore almost doomed to failure. The
$b$-expansion will on the other hand be much more stable because at
each order in $b$ one will effectively be able to resum
collinearly-enhanced terms $\as^n(\as Y)^m$, in analogy with what is
done in \cite{CCS2,GAVINRESUM}. The detailed behaviour is currently under
investigation \cite{CCSS3}.

It is important keep in mind that after accounting for higher-order
effects, the numerical values in figures such as Fig.~\ref{fig:ymax}
will be significantly altered, because on the one hand the diffusion
coefficient is known to be very different at higher orders
\cite{NLLBFKL}, and the tunneling behaviour is also expected to be
numerically substantially modified. Nevertheless, we expect the
qualitative features of that figure to persist.

Finally, one potential practical limitation of the $b$ expansion is
the case when $t$ and $t_0$ are  different, leading to
the need for a collinear resummation. In such a case, as mentioned in 
Sec.3.2,  in the regime where
$t-t_0 \ll t_0$,  the coefficients
of the $b$ expansion will be enhanced by powers of $(t-t_0)$, and
though the series may still be formally convergent, it is not clear
whether it will be practical to include a sufficient number of terms
for the convergence to be reached. To establish this point requires
further investigation.

\section{Conclusions}

In this paper we have studied the properties of the gluon Green's
function in the case of the small-$x$ evolution equation with
running coupling.  In general the solution can be decomposed into
perturbative --- hard Pomeron and non-perturbative --- Pomeron
components.
The hard Pomeron is then governed by the perturbatively calculable
saddle point exponent $\oms(t)$ which is modified by the corrections
due to running coupling effects.  The non-perturbative part has
instead a true singularity $\omp$ which is leading at large values of
rapidity since $\omp > \oms(t)$.  The Pomeron part is however
suppressed at large values of $t$ by a tunneling factor which has a
universal form of $\exp[-1/(b \asb) g(\asb)]$. As a simple
illustration, in the Airy diffusion model the Pomeron singularity is
the lowest lying energy state of the potential in the
Schroedinger-like problem. The perturbative part corresponds then to
the continuum 
of states at $\mathrm{Re} \, \om <0$. The decomposition into perturbative
and non-perturbative parts is however non unique in the sense that
some subleading bound states $\om_b < \omp$ can be included in the
perturbative part of the gluon Green's function and are similar to
a renormalon ambiguity.

Starting from this qualitative picture, we introduced the $b
\rightarrow 0 $, $\asb$ fixed limit in which the Pomeron is
suppressed, in order to study the properties of the perturbative part
$G_{pert}(Y;t,t_0)$.  By applying the $b$ expansion we obtained the
saddle point exponent and the diffusion corrections.  We showed that
apart from the well known $\sim \asb^5 Y^3$ corrections there are
further subleading ones of type $\sim Y^2$ and $\sim Y$.  The latter
constitutes a second order $\oms$ shift, which adds up to dynamical
corrections of the same order $b^2 \asb^2 Y$ coming from subleading
corrections to the kernel. We have investigated in particular a scale
change in the running coupling \cite{CCSS3}.  We found out that this
perturbative expansion is controlled by the parameter
$z=\frac{1}{2}(b\asb)^2 (\oms Y)^2 \chi_m^{''}$ and is valid for $z
\ll 1$. Outside this regime, when $z \gtrsim 1$, the perturbative
solution starts to oscillate and the non-perturbative Pomeron takes
over.  Thus the genuine hard Pomeron part of the gluon Green's
function which is governed by $\oms(t)$ can be studied
phenomenologically at large values of $t$ and moderate rapidities, in
the limited regime where the non-perturbative part is strongly
suppressed.

Additional NLL corrections are going to change the picture of
diffusion and tunneling in a quantitative way. For example the so
called kinematical constraint \cite{KC}, incorporated in some
resummations of subleading corrections \cite{CCSS3} can be shown to
increase the rapidity where tunneling occurs by about
$t-t_0$; the proportionately larger effect of higher-order corrections
at larger values of $\as$ will also delay the onset of tunneling, by
reducing the non-perturbative Pomeron exponent. Both these effects
can be expected to widen the window for phenomenological tests of the
purely perturbative hard Pomeron.
It should be also remarked that unitarity effects can in principle
change significantly the phenomenon of tunneling. It was noticed in
\cite{Balitsky2,GBMS} that in the case of the non-linear small $x$
evolution equation \cite{Balitsky1,Kovchegov}, the generation of the
saturation scale $Q_s(x)$ leads to the suppression of diffusion into
the low scales $k<Q_s(x)$ and the distribution of the momenta is
driven towards the perturbative regime.  More detailed theoretical
studies of these phenomena are thus needed.

\section{Acknowledgments}
This work was supported in part by the E.U. QCDNET contract
FMRX-CT98-0194, MURST (Italy), 
the  Alexander von Humboldt Stiftung and the  Polish Committee for
Scientific Research (KBN) grants no. 2P03B 05119, 2P03B 12019, 5P03B
14420.


\renewcommand{\theequation}{A.\arabic{equation}}
\setcounter{equation}{0}  
\section*{Appendix A: Calculation of the fluctuations}

The fluctuation matrix is obtained by taking the second derivatives
with respect to $\gamma,\delta$ and $\om$ of the exponent 
\begin{equation}
E(\gamma,\delta,\omega) = \om Y + (\gamma - 1/2) t -(
\gamma+\delta-1/2) t_0 - \frac{1}{b\om} X(\gamma) + \frac{1}{b\om}
X(\gamma+\delta)
\end{equation}
in the double-$\gamma$ representation Eq.~(\ref{greenpertgam}).
It reads as follows:
\begin{equation}
\frac{1}{b\omb} \left(\begin{array}{ccc}
-2 \chip & -  \chip & 0 \\
& & \\
- \chip & -  \chip & - bt \\
& & \\
0 & - bt & 2 b Y \\
\end{array} \right)
\end{equation}
where we have made use of  the relations (\ref{delta}).
The secular equation reads then
\begin{eqnarray}
\lefteqn{\left(\frac{2\chip}{b \omb} + \lambda \right)
  \left(\frac{\chip}{b \omb} + \lambda \right) \left(\frac{2 Y}{\omb}
    -\lambda \right) + } \nonumber \\
& +  &  \left(\frac{2\chip}{b\omb} + \lambda \right) \left(
  \frac{t}{\omb}\right)^2 + \left(\frac{\chip}{b\omb} \right)^2
\left(\lambda - \frac{2 Y}{\omb} \right) \;  = \; 0  \; .
\label{eigenvalue}
\end{eqnarray}
By looking at the separate terms in Eq.~(\ref{eigenvalue}) it turns out
that the sum and product of the eigenvalues have the following form
($\chip<0$)
\begin{eqnarray}
\lambda_1 + \lambda_2 + \lambda_3 &  = & -\frac{3\chip}{b\omb} +
\frac{2 Y}{\omb} >0 \nonumber \\ 
\lambda_1 \lambda_2 \lambda_3 & = & \frac{2 \chip}{ (b \omb)^3} [
(bt)^2 + b Y \chip] <0 \; ,
\label{sumproduct}
\end{eqnarray}
which means that we have two positive eigenvalues and one negative -
provided that 
\begin{equation} 
  (b\,t)^2 > -\chip\, b \, Y \; .
\label{cond}
\end{equation}
Since we have $b Y = \zeta (bt)^2 /\chi_m$, the condition
(\ref{cond}) is equivalent to $\chi_m > -\zeta \chip$ and holds
as long as $\zeta$ is a small parameter. Two positive eigenvalues
correspond thus to fluctuations along the two imaginary axes and the
negative eigenvalue to the fluctuation along the real axis in
Eq.~(\ref{greenpertgam}). After performing the integrations we obtain
the following overall normalisation factor (using (\ref{sumproduct}))
\begin{equation} 
  N(t,Y) \; = \; \frac{1}{(2\pi)^2} \frac{t_0}{\omb} (2 \pi)^{3/2}
\frac{1}{\sqrt{-\lambda_1 \lambda_2 \lambda_3}} =
\sqrt{\frac{b\omb}{-4 \pi \chi^{\prime}(\bar{\gamma})} }
\frac{1}{\sqrt{1 + \frac{ \chi^{\prime}(\bar{\gamma})}{\chi_m }\zeta}}
\; .
\label{normfactora}
\end{equation}
From (\ref{normfactora}) we can derive the subleading $\sim Y^2$
corrections.  To this aim we expand, using (\ref{expansions}), $\omb$
and $\chip$ in $\zeta$ as follows
\begin{eqnarray}
\omb & = & \oms(t) (1 + D \frac{\zeta^2}{4}) + \dots \nonumber \\ 
\chi^{\prime}(\gamb=\frac{1}{2}-\bar{p}) & = & -\bar{p} \,
\chi^{\prime \prime}(1/2) - \frac{1}{3!} \, \bar{p}^3 \,
\chi^{(4)}(1/2) = \nonumber \\ 
& = &  -  \frac{\zeta}{2}(1+\frac{5}{12}D \zeta^2) \left[
  \chi^{\prime \prime}(1/2)+ \frac{1}{3!} \left(\frac{\zeta}{2}
  \right)^2  \chi^{(4)}(1/2) \right] + \dots \; ,
\label{omchiexp} 
\end{eqnarray}
with $D=\chi_m^{''}/(2\chi_m)$.  Inserting (\ref{omchiexp}) into
(\ref{normfactora}) gives
\begin{eqnarray}
N(t,Y) & = & \sqrt{\frac{b\oms}{2 \pi \zeta \chi_m^{''}}}
(1+\frac{1}{8}D\zeta^2)(1-\frac{5}{24} D \zeta^2) (1-\frac{1}{48}
\frac{\chi_m^{(4)}}{\chi_m^{''}}) (1+\frac{1}{2} D \zeta^2) =
\nonumber \\
& = & \sqrt{\frac{1}{4 \pi D \oms Y}} \left[1 + D \zeta^2
  \left(\frac{5}{12} - \frac{1}{24} \frac{\chi_m^{(4)}
      \chi_m}{(\chi_m^{''})^2} \right) \right] \; .
\label{normfactorapp}
\end{eqnarray}
\renewcommand{\theequation}{B.\arabic{equation}}
  \setcounter{equation}{0}  
\section*{Appendix B: Calculation of $Y^2$ and $Y$ terms in the Airy
  approximation} 

We show here how to obtain the quadratic $\sim Y^2$ and linear $\sim
Y$ corrections to the saddle point exponent $\oms(t)$ in the case of
the Airy diffusion model. We start from the exponent in the expression
(\ref{gpertsolz})
\begin{equation}
E \; = \; \oms(\tilde{t}) |Y| \bigg[\frac{\xi(\xi - 1)^2}{(-z)/3} +
\bigg(\frac{\Delta t}{2\tilde{t}}\bigg)^2 \frac{\xi}{(-z)} +
\frac{1}{\xi}\bigg] \; ,
\label{exponentai}
\end{equation}
and consider the expansion around the saddle point $\bar{\xi}$
defined by equation (\ref{zfun}).  We obtain
\begin{equation}
E = a_0 + a_1 \Delta \xi + \frac{1}{2} a_2 (\Delta \xi)^2 +
\frac{1}{3!} a_3 (\Delta \xi)^3 + \dots \; ,
\end{equation}
where we defined
\begin{equation}
a_n \equiv {\partial^n E\over \partial \xi^n }
\bigg|_{\xi=\bar{\xi}} \; ,
\end{equation}
and
\begin{equation}
\Delta \xi \equiv \xi - \bar{\xi} \; .
\end{equation}
For simplicity we take ${\Delta t \over 2 \tilde{t} }= 0$.  The saddle
point condition $a_1=0$ results in equation (\ref{zfun}) which at
small values of $z$ has a solution (\ref{leadsaddle})
$\bar{\xi}=1-\frac{1}{6}z$ .  The term $a_0$ instead provides the
leading exponent $\oms$.  The second derivative evaluated at
$\bar{\xi}$ reads then
\begin{equation}
a_2  \; = \; 2 \oms Y \left[{-3(3 \bar{\xi} -2) \over z} + {1 \over
    {\bar{\xi}}^2 }\right] \; ,
\end{equation} 
and the third one 
\begin{equation}
a_3  \; = \; - 6 \oms Y  \left[{3 \bar{\xi} \over z} +
  \frac{1}{\bar{\xi}^4} \right] \; .
\end{equation}
We then expand  the exponential in the following form
\begin{equation}
e^{\frac{1}{2} a_2 (\Delta \xi)^2 + \frac{1}{3!} a_3 (\Delta
  \xi)^3} \simeq 
e^{\frac{1}{2} a_2 (\Delta \xi)^2} \,[ 1 + \frac{1}{3!} a_3 (\Delta
\xi)^3 + \frac{1}{2} \left(\frac{a_3}{3!}\right)^2 (\Delta \xi)^6
\, ] \; .
\label{expandexp}
\end{equation}

By integrating over $\Delta \xi$ the expression (\ref{expandexp}) we
arrive at the following result
\begin{equation}
\sqrt{\frac{2\pi}{a_2}} \left(1  +\frac{15}{16} \frac{2^3
    a_3^2}{(3!)^2 a_2^3} \right) \; .
\label{fluct1}
\end{equation}

The first term in (\ref{fluct1}) i.e. the factor $\sqrt{2\pi/a_2}$
comes from the second order fluctuation and together with
normalisation in (\ref{gpertsolz}) gives the following overall
normalisation factor to $G_{pert}(Y;t,t_0)$ and the $Y^2$ term
\begin{eqnarray}
\frac{1}{2\pi \sqrt{D|z|/3}}  \sqrt{\frac{\pi}{\oms Y \left[{-3 \over
        z} (3 \bar{\xi} -2)  + {1 \over {\bar{\xi}}^2 }\right]}} &
\simeq & 
 \frac{1}{\sqrt{4 \pi D \oms Y }} \frac{1}{\sqrt{1-\frac{5}{6}z}}
 \simeq \nonumber \\
& \simeq & \frac{1}{\sqrt{4 \pi D \oms Y }} (1 + \frac{5}{12} z) \; ,
\label{fluct2}
\end{eqnarray}
which checks with the first term $\sim D\zeta^2$ in
Eq.~(\ref{normfactorapp}) (due to the Airy approximation we will not
reproduce the second term with higher order derivatives).

The second term provides  in turn the linear term  in $Y$
\begin{equation}
\frac{15}{16} \frac{2^3 a_3^2}{(3!)^2 a_2^3} \simeq
\frac{15}{16}\left(\frac{3 \oms Y}{z}\right)^2 \left(\frac{z}{3 \oms
    Y}\right)^3 = \frac{5}{16} \frac{z}{\oms Y} = \frac{5}{32}
(b\asb)^2 \asb \chi_m^{\prime\prime} Y \; ,
\end{equation}
which is a second order shift to the saddle point value $\oms(t)$.
The $(\Delta \xi)^4$ term in (\ref{expandexp}) can be safely
neglected since it is of the subleading order $(b \asb)^4$.



\end{document}